\documentclass[12pt]{iopart}

\usepackage{amssymb}
\usepackage{hyperref}
\usepackage{esint}
\usepackage{cite}
\usepackage[disable]{todonotes}

\begin{document}

\title{The Quench Action}

\author{Jean-S\'ebastien Caux}

\address{Institute for Theoretical Physics, University of Amsterdam, Science Park 904, 1098 XH Amsterdam, The Netherlands}

\begin{abstract}
We give a pedagogical introduction to the methodology of the Quench Action, which is an effective representation for the calculation of time-dependent expectation values of physical operators following a generic out-of-equilibrium state preparation protocol (for example a quantum quench). The representation, originally introduced in \cite{2013_Caux_PRL_110}, is founded on a mixture of exact data for overlaps together with variational reasonings. It is argued to be quite generally valid and thermodynamically exact for arbitrary times after the quench (from short times all the way up to the steady state), and applicable to a wide class of physically relevant observables. Here, we introduce the method and its language, give an overview of some recent results, suggest a roadmap and offer some perspectives on possible future research directions.
\end{abstract}

\maketitle
\tableofcontents

\listoftodos

\newpage

\section{Introduction}
There has recently been a lot of interest in the study of relaxation in isolated many-body quantum systems (see \cite{2011_Polkovnikov_RMP_83,2015_Eisert_NP_11,2016_Gogolin_RPP_79,arXiv150906411} and references therein), fuelled in no small part by experimental studies of out-of-equilibrium situations using cold atoms 
\cite{2002_Greiner_NATURE_419,2006_Kinoshita_NATURE_440,2007_Hofferberth_NATURE_449,2012_Trotzky_NATPHYS_8,2012_Cheneau_NATURE_481,2012_Gring_SCIENCE_337,2013_Kuhnert_PRL_110,2014_Meinert_SCIENCE_344,2015_Langen_SCIENCE_348}. Theoretically, the problem boils down to considering a generic initial state, which we assume to be a pure state represented by a wavefunction $|\Psi (t=0)\rangle$, evolving unitarily in time according to a Hamiltonian $H$, 
\begin{equation}
|\Psi (t) \rangle = e^{-i H t} |\Psi (t=0) \rangle,
\end{equation}
where $|\Psi (t=0) \rangle$ is not a simple ({\it e.g.} finite, zero entropy in the thermodynamic limit) 
linear combination of eigenstates of $H$. Such initial states can be constructed (among other ways) by
performing a sudden quantum quench \cite{2006_Calabrese_PRL_96,2007_Calabrese_JSTAT_P06008}, {\it i.e.} preparing the initial state as an eigenstate of a certain Hamiltonian $H_{<}$
but suddenly changing the latter to $H$ from $t=0$ onwards.

For many reasons, the one-dimensional case turns out to be of particular interest, among others due to the existence of exact methods (often resting on the property of integrability) allowing to perform nonperturbative computations. One of the central developments of recent years has been the realization that time evolution under an integrable $H$ must be viewed as sitting in a different class as compared to the non-integrable case. In the absence of integrability, relaxation to a thermal ensemble is expected to occur; instead, for the integrable case, thermalization should not occur, and the expectation values of operators long after the quench are expected to be described by a generalized Gibbs ensemble (GGE) \cite{2007_Rigol_PRL_98,2008_Rigol_NATURE_452} (see also the accompanying paper \cite{arXiv1604.03990}), taking not only the energy conservation constraint but also those associated to all other available conserved charges into account. 

Despite its appeal, the main drawback of the GGE is its limitation to the description of long-time averages only. The description of the whole time evolution, from small to large times, is however desirable for quantitative experimental phenomenology, in which the long time limit can be difficult to reach. In any case, the time evolution process is itself extremely interesting to study, since it encodes how quantum interference effects can eventually drive this out-of-equilibrium initial state to a well-defined steady state. 

Our main focus will be the calculation of post-quench time-dependent expectation values of generic (combinations of) physical operators ${\cal O}$, 
\begin{equation}
\bar{\cal O} (t) \equiv \frac{ \langle \Psi (t) | {\cal O} | \Psi (t) \rangle}{ \langle \Psi (t) | \Psi (t) \rangle}.
\label{eq:Obart}
\end{equation}
This problem is particularly difficult not only because the time evolution of the state itself is highly nontrivial, but also because the operators ${\cal O}$ one is typically interested in create a complicated web of excitations when acting on a given state. 

A new take on this problem, inspired by the generic usefulness of variational reasonings in physics, was recently proposed in \cite{2013_Caux_PRL_110}, allowing to transform (\ref{eq:Obart}) into a much more efficient and (in principle) easily implementable representation, valid for situations in which energy and higher charge fluctuations around the steady state become small in the thermodynamic limit. The approach is based on the definition of an appropriate measure in Hilbert space, the Quench Action, aiming at quantifying the relative importance of eigenstates in the time evolution. The approach has a number of benefits: it can, as the discussed examples demonstrate, be implemented exactly in the thermodynamic limit for quenches to nontrivially-interacting theories, and most importantly it offers a very clear path towards the calculation of the actual time dependence of observables, and not just on the theoretically most easily accessible, but experimentally much less easily obtainable steady state. The method thus invites further applications with a slight change of focus away from the $t\rightarrow \infty$ limit and towards finite time scales which anyway carry even richer physics.

The paper is organized as follows. We start in section \ref{sec:Fund} by recalling important features of integrable Hamiltonians, and defining some useful notations. The approach towards the thermodynamic limit is then presented, with a discussion of a handy implementation of the resolution of the identity for the problem at hand, followed by a summary of important notions of the Algebraic Bethe Ansatz. In section \ref{sec:QA}, the Quench Action formalism is then introduced for the study of a generic quench problem, keeping a focus on the calculation of time-dependent expectation values, for which a greatly simplified representation is given. Section \ref{sec:Tour} presents a quick summary of recently worked-out solutions to quench problems using the method. This is followed in section \ref{sec:Road} by some perspectives on the general problem of computing overlaps between initial states and exact eigenstates, these overlaps being the fundamental building blocks on which the method is based, and with some further perspectives on potential further developments. 

\section{Fundamentals}
\label{sec:Fund}
Let us begin by defining the language and notations we will use later on. 

\subsection{Bethe states}
Upon quantization, the eigenstates of a Bethe Ansatz-solvable model \cite{GaudinBOOK,KorepinBOOK,TakahashiBOOK,HubbardBOOK} are uniquely labeled by a set of quantum numbers $\{ I \}$. For simplicity, we consider a model in which only one type of particle is present (the generalization to cases where multiple types of string states are possible, {\it e.g.} for spin chains (see discussion around equation (\ref{eq:BAEXXZstrings})), is completely straightforward, in which case the quantum numbers carry a `particle type' index). If the model is in the continuum, the possible quantum numbers are unbounded; on the lattice, they obtain type-specific limits; we do not specify these model-dependent details at this stage. A resolution of the identity operator in Fock space can then be formally written as\footnote{The details of how to do this vary for each model, and can include some (surmountable) difficulties such as dealing with global symmetries, `beyond-the-equator' states (see {\it e.g.} \cite{2002_Baxter_JSP_108}) etc.}
\begin{equation}
{\bf 1} = \sum_{\{ I \}} | \{ I \} \rangle \langle \{ I \}| = \sum_N \sum_{\{ I \}_N} | \{ I \}_N \rangle \langle \{ I \}_N|
\end{equation}
in which the summation over $N$ (the number of particles) runs over all allowed values. We use the bra-ket notation for eigenstates, 
namely $| \{ I \}_N \rangle$ represents the normalized $N$-particle Bethe Ansatz state
\begin{equation}
| \{ I \}_N \rangle \equiv {\cal N} (\{ I \}_N) \prod_{j=1}^N B(\lambda_j(\{ I \}_N)) | 0 \rangle
\label{eq:states}
\end{equation}
obtained by acting on the reference state $|0\rangle$ with raising operators $B(\lambda)$ (see subsection \ref{subsec:BA}) evaluated at the solutions of the Bethe equations in terms of rapidities $\lambda_j$ corresponding to quantum numbers $\{ I\}_N$,
\begin{equation}
\theta_{kin} (\lambda_j) + \frac{1}{L} \sum_{l=1}^N \theta_{scat} (\lambda_j - \lambda_l) = \frac{2\pi}{L} I_j, \hspace{2cm} j = 1, ..., N,
\label{eq:BE}
\end{equation}
where $L$ is the system size in dimensionless units and $\theta_{kin}$ and $\theta_{scat}$ are model-dependent kinetic and scattering kernels. 
In (\ref{eq:states}), ${\cal N} (\{ I \}_N)$ is the state normalization factor obtained from the appropriate Gaudin determinant. For definiteness, in (\ref{eq:states}), we have explicitly written each rapidity as being a function of the whole set of quantum numbers. This fact, which the reader should not forget, will be kept implicit in the following.

\subsection{The thermodynamic limit: root distributions and root densities}
The general idea behind the thermodynamic limit is (as far as individual states are concerned) to replace sets of rapidities by distribution functions, and (and as far as state ensembles are concerned) to express the summations over eigenstates by a form of functional integration, with an eye towards the obtention of analytical results. The first step in this direction is to think of the quantum numbers scaled by the system size $I_j/L$ in terms of a real variable $x$ taking values on the real line. We associate to each eigenstate a root distribution $\rho_{dist}$ with support on the real axis
\begin{equation}
|\{ I \} \rangle \Leftrightarrow \rho_{dist} (x; \{ I \}),
\end{equation}
this distribution in $x$ space being defined as
\begin{equation}
\rho_{dist} (x; \{ I \}) = \frac{1}{L} \sum_j \delta (x - x_j), \hspace{2cm} x_j \equiv \frac{I_j}{L}.
\label{eq:rhodistdef}
\end{equation}
Of course, the particular system being studied might be such that the interval of support of $x$ is bounded; in all cases however these bounds are state-independent and their presence does not affect the reasonings presented here.

It is important to remember that $\rho_{dist}$ carries the complete characteristics of a given eigenstate, and is therefore (implicitly) labeled by the set $\{I \}$. Note also that it is not a continuously differentiable function of $x$, but rather a distribution visualized as an unevenly-spaced but equal-height delta comb defined by the quantum numbers. It exactly (to all orders in $1/L$) obeys the identities
\begin{equation}
\int dx  \rho_{dist} (x; \{ I \}_N) = \frac{N}{L}, \hspace{1cm}
\int dx \rho_{dist} (x; \{ I \}_N) f(x) = \frac{1}{L} \sum_{j=1}^N f(x_j)
\end{equation}
in which the integral is taken over the whole support of $\rho(x)$ and $f(x)$ is any continuous function.

The thermodynamic limit, denoted $\lim_{Th}$ in the following, is defined as the simultaneous limits $L \rightarrow \infty$, $N \rightarrow \infty$ with $N/L$ fixed. For later purposes, we will also consider a more flexible viewpoint here and split the space of quantum numbers into `boxes' (labeled by an integer $i$) of size $l_i$ chosen such that $\lim_{Th} \frac{l_i}{L} = 0$ and $\lim_{Th} l_i = +\infty$ $\forall i$, giving us a `box regularized' thermodynamic limit which we will denote by $\lim_{Th,reg}$. 

For a given eigenstate, within box $i$, there will be a positive integer number $n_i$ of occupied quantum numbers, allowing to define an average density $\rho_i$ in box $i$ as
\begin{equation}
\rho_i = \frac{n_i}{l_i}, \hspace{1cm} \mbox{with} \hspace{1cm} 0 \leq \rho_i \leq 1 ~\forall i.
\label{eq:rhoi}
\end{equation} 
The specification of the box fillings $\rho_i$ does not label eigenstates uniquely. In order to do so, we also need to specify the particular configuration of quantum numbers within all boxes. Assuming that $c_i$ (which is an index running over ${\tiny \left( \begin{array}{c} l_i \\ n_i \end{array} \right)}$ values) labels such a configuration in box $i$, we can thus uniquely label an eigenstate for a given box regularization by its set of density and in-box configuration labels, $| \{ I \} \rangle \equiv | \{ \rho_i \}; \{ c_i \} \rangle$, the resolution of the identity becoming
\begin{equation}
{\bf 1} = \sum_{\{ I \} } | \{ I \} \rangle \langle \{ I \} | = \sum_{\{ \rho_i \}} \sum_{\{ c_i \}} | \{ \rho_i \}; \{ c_i \} \rangle \langle \{ \rho_i \}; \{ c_i \} |.
\label{eq:ib1}
\end{equation}

In the Hilbert space, for a given box regularization, there will be many states with the same set of box densities $\{ \rho_i \}$. This number is simply
\begin{equation}
\sum_{\{ c_i \} } \left. 1 \right|_{\small \mbox{fixed}~ \{ \rho_i \}} = \prod_i \left( \begin{array}{c} l_i \\ n_i \end{array} \right) \equiv e^{S^{YY}_{\{ \rho_i \}}}
\end{equation}
whose logarithm has a leading extensive term which is the well-known Yang-Yang entropy \cite{1969_Yang_JMP_10}.

Within the set of states at fixed $\{ \rho_i \}$, there exists one with a `maximally flat' configuration where, in all boxes $i$, the $n_i$ occupied quantum numbers are distributed uniformly over the $l_i$ available spaces\footnote{The precise details obviously don't really matter here.}. We will label this as $\{ c_i \} = \{ 0 \}$. We can then write
\begin{equation}
\rho_{dist} (x; \{ I \}) = \rho_{dist} (x; \{ \rho_i \}; \{ c_i \})
= \rho (x; \{ \rho_i \}) + \rho_{ib} (x; \{ \rho_i \}; \{ c_i \})
\end{equation}
where
\begin{eqnarray}
\rho (x; \{ \rho_i \}) \equiv \rho_{dist} (x; \{ \rho_i \}; \{ 0 \}), \nonumber \\
\rho_{ib} (x; \{ \rho_i \}; \{ c_i \}) \equiv \rho_{dist} (x; \{ \rho_i \}; \{ c_i \}) - \rho (x; \{ \rho_i \})
\end{eqnarray}
respectively represent the `maximally flat' part of the distribution, and its in-box microscopic features.
In the regularized thermodynamic limit, we can effectively make the replacements
\begin{equation}
\lim_{Th,reg} \rho (x; \{ \rho_i \}) = \rho_{sm} (x) \in C^\infty, \hspace{10mm}
\lim_{Th,reg} \rho_{ib} (x; \{ \rho_i \}; \{ c_i \}) = 0
\label{eq:rhosm}
\end{equation}
as far as any in-box-configuration insensitive quantity is concerned, 
in the sense that given a continuously differentiable function $f(x)$, we have
\begin{eqnarray}
\lim_{Th,reg} \int dx \rho (x; \{ \rho_i \}) f(x) = \int dx \rho_{sm} (x) f(x), 
\nonumber \\
\lim_{Th,reg} \int dx \rho_{ib} (x; \{ \rho_i \}; \{ c_i \}) f(x) = 0.
\label{eq:rhosm2}
\end{eqnarray}
The smooth, differentiable function $\rho_{sm}(x)$ will be called the {\it root density} in order to distinguish it from the neither smooth nor differentiable distribution $\rho_{dist}$. Another convenient function to introduce is the hole density function, which is the complement of $\rho_{dist}$ in the space of quantum numbers, and the total density $\rho^t = \rho + \rho^h$ whose smooth version coincides with unity,
\begin{equation}
\rho^t_{sm} (x) = \rho_{sm}(x) + \rho^h_{sm}(x) = 1.
\end{equation}

The notations up to now have involved integrations over quantum number space. It is customary to also express integrals in rapidity space, extending the Bethe equations (\ref{eq:BE}) to the functional equation
\begin{equation}
\theta_{kin} (\lambda) + \int d\lambda' ~\theta_{scat} (\lambda - \lambda') \rho_{sm}(\lambda') = 2\pi ~x (\lambda)
\label{eq:BEext}
\end{equation}
where the function $x(\lambda)$ is now viewed as labelling the state. This function then defines a mapping between the quantum number and rapidity spaces. The transformation rule of $\delta$ functions then allows to write the densities directly in rapidity space,
\begin{equation}
\rho_{sm}(\lambda) = \rho_{sm}(x (\lambda)) \frac{d x(\lambda)}{d\lambda}, \hspace{20mm} \rho^t_{sm} (\lambda) = \frac{d x(\lambda)}{d\lambda}.
\label{eq:rhosdefs}
\end{equation}
A technical aspect to bear in mind is that the distributions so defined are not made of equal-height delta functions, but rather of state-dependent heights, which are implicit functions of the interaction-induced backflows when moving rapidities around. The quantum numbers are the truly statistically-independent parameters from which the trace over states should be defined. 

\subsection{Rudiments of the Algebraic Bethe Ansatz}
\label{subsec:BA}
For completeness, we recall a few basic notions of the Algebraic Bethe Ansatz which are useful for the discussions here. Readers looking for a more systematic introduction to the subject should consult \cite{KorepinBOOK} and references therein.

At the center of the ABA is the notion of the $R$-matrix, which is a matrix defined in the tensor product of two auxiliary spaces ${\cal A}_1 \otimes {\cal A}_2$, obeying the Yang-Baxter relation
\begin{equation}
R_{12} (\lambda,\mu) R_{13}(\lambda,\nu) R_{23}(\mu,\nu) = R_{23} (\mu,\nu) R_{13}(\lambda,\nu) R_{12}(\lambda,\mu).
\label{eq:YB}
\end{equation}
One then introduces a monodromy matrix $T(\lambda)$ living in the tensor product of auxiliary and Hilbert spaces and depending on a (in general complex-valued) spectral parameter $\lambda$, and obeying the intertwining relation 
\begin{equation}
R_{12} (\lambda, \mu) T_1 (\lambda) T_2 (\mu) = T_2 (\mu) T_1 (\lambda) R_{12} (\lambda, \mu).
\label{eq:RTTequalsTTR}
\end{equation}
In the simplest cases, the auxiliary space is two-dimensional and the monodromy matrix is represented as  
\begin{equation}
T (\lambda) \equiv \left( \begin{array}{cc}
A (\lambda) & B(\lambda) \\
C (\lambda) & D(\lambda)
\end{array} \right)
\label{eq:TisABCD}
\end{equation}
in which $A(\lambda), B(\lambda), C(\lambda), D(\lambda)$ are operators acting in Hilbert space. The transfer matrix $\tau$, defined as the trace (in auxiliary space) of the monodromy matrix (the second equality here is special to the case of two-dimensional auxiliary space)
\begin{equation}
\tau(\lambda) = \mbox{Tr}_{\cal A} T(\lambda) = A(\lambda) + D(\lambda)
\label{eq:tauistraceT}
\end{equation}
then has the fundamental property of self-commutation at arbitrarily chosen spectral parameters,
\begin{equation}
\left[ \tau(\lambda), \tau(\mu) \right] = 0, ~~ \forall ~\lambda, \mu ~\in ~\mathbb{C}.
\label{eq:taucommutes}
\end{equation}
This naturally allows to build a set of commuting quantum charges from so-called trace identities
\begin{equation}
Q_n \propto \frac{d^n}{d\lambda^n} \ln \tau (\lambda) |_{\lambda = \xi}
\label{eq:Infromtau}
\end{equation}
in which the constant prefactors can be chosen at leisure, and $\xi$ is an evaluation point chosen to make the charges local (the $R$-matrix then becoming of permutation form).

All fundamental commutation relations of the model are contained in the algebra of monodromy matrix operators $A, B, C, D$ issuing from (\ref{eq:RTTequalsTTR}) once an $R$-matrix and a monodromy matrix have been defined. The algebra is quadratic in operators. For example, the simplest nontrivial structure for an $R$-matrix takes the shape
\begin{equation}
R (\lambda, \mu) = \left( \begin{array}{cccc} 
    1 & 0 & 0 & 0 \\
    0 & b (\lambda, \mu) & c (\lambda, \mu) & 0 \\
    0 & c (\lambda, \mu) & b (\lambda, \mu) & 0 \\
    0 & 0 & 0 & 1
\end{array} \right)
\label{eq:R1}
\end{equation}
with some constraints (which we do not write down here) on the $b,c$ functions to ensure algebraic consistency through the Yang-Baxter relation. The intertwining relations (\ref{eq:RTTequalsTTR}) then imply
\begin{eqnarray}
\left[ A(\lambda), A(\mu) \right] = 0, \label{eq:AA} \\
\left[ B(\lambda), B(\mu) \right] = 0, \label{eq:BB} \\
A(\lambda) B(\mu) = \frac{1}{b(\mu, \lambda)} B(\mu) A(\lambda) - \frac{c(\mu, \lambda)}{b(\mu, \lambda)} B(\lambda) A(\mu), \label{eq:AB}\\
\left[ B(\lambda), C(\mu) \right] = \frac{c(\lambda,\mu)}{b(\lambda,\mu)} ( D(\mu) A(\lambda) - D(\lambda) A(\mu)), \label{eq:BC}
\end{eqnarray}
and 12 other similar relations which we leave out here for the sake of brevity (see {\it e.g.} \cite{KorepinBOOK} for additional details).

Assuming the existence of a reference state $|0\rangle$ such that 
\begin{equation}
A(\lambda) | 0 \rangle = a(\lambda) | 0 \rangle, \hspace{1cm} D(\lambda) | 0 \rangle = d(\lambda) | 0 \rangle, \hspace{1cm}
C(\lambda) | 0 \rangle = 0
\label{eq:vacuum}
\end{equation}
where the functions $a(\lambda)$ and $d(\lambda)$ can be chosen arbitrarily (they should be viewed as model-defining choices), states in Hilbert space can be constructed by acting with the `raising' operators $B(\lambda)$ according to 
\begin{equation}
| \{ \lambda_j \}_M \rangle \equiv \prod_{j=1}^M B (\lambda_j) | 0 \rangle
\label{eq:Bonvacuum}
\end{equation}
for generic $M$ and $\{ \lambda_j \}_M$.  Note the very important fact that the order in the product is immaterial,
in view of the commutation relation (\ref{eq:BB}). So written, the states are also not normalized. Again using the monodromy matrix element commutation relations, one can show that these states are eigenstates of the transfer matrix (\ref{eq:tauistraceT}) provided the set of rapidities $\{ \lambda \}$ obeys the Bethe equations
\begin{equation}
\frac{a(\lambda_j)}{d(\lambda_j)} \prod_{l \neq j} \frac{b(\lambda_j, \lambda_l)}{b(\lambda_l, \lambda_j)} = 1 ~~~ \forall ~j = 1, ..., M,
\label{eq:BER1}
\end{equation}
whose logarithm (allowing the introduction of quantum numbers $\{ I\}$, which allow to classify eigenstates once the Pauli principle of non-coincident rapidities has been implemented) we have written as (\ref{eq:BE}). These states diagonalize the transfer matrix
\begin{equation}
\tau (\lambda) | \{ \lambda_j \}_M \rangle = \tau (\lambda | \{ \lambda_j \}_M ) | \{ \lambda_j \}_M \rangle,
\label{eq:taueigenstates}
\end{equation}
with eigenvalues
\begin{equation}
\tau (\lambda | \{ \lambda_j \}_M ) = a(\lambda) \prod_{j=1}^M b^{-1} (\lambda_j, \lambda) + d(\lambda) \prod_{j=1}^M b^{-1} (\lambda, \lambda_j),
\label{eq:taueigenvalues}
\end{equation}
this defining the eigenvalue of all charges (\ref{eq:Infromtau}).

A further pillar of the Algebraic Bethe Ansatz is Slavnov's theorem \cite{1989_Slavnov_TMP_79}, which gives the overlap between states
\begin{eqnarray}
S_M (\{\mu \}, \{ \lambda \}) = \langle 0 | \prod_{j=1}^M C(\mu_j) \prod_{k=1}^M B(\lambda_k) | 0 \rangle
\label{scalar_prod}
\end{eqnarray}
as a computationally convenient determinant, provided {\it either} $\{ \lambda \}$ {\it or} $\{ \mu \}$ obeys the Bethe equations:
{ \setlength{\mathindent}{10mm}
\begin{eqnarray}
S_M (\{ \mu \}, \{ \lambda \}) = \frac{\prod_{j=1}^M \prod_{k=1}^M \varphi (\mu_j - \lambda_k)}{\prod_{j < k} \varphi (\mu_j - \mu_k)
\prod_{j > k} \varphi(\lambda_j - \lambda_k)} \det T(\{ \mu \}, \{ \lambda \}), 
\label{Slavnov}
\end{eqnarray}
}\noindent
where
\begin{equation}
T_{ab} = \frac{\partial}{\partial \lambda_a} \tau (\mu_b| \{ \lambda \})
\end{equation}
and $\varphi$ is a model-dependent scalar function.
The importance of Slavnov's theorem cannot be overemphasized, since it allows for the computation of matrix elements in integrable models once the solution to the `quantum inverse problem' (the mapping of physical operators to monodromy matrix operators, \cite{q-alg/9612012,1999_Kitanine_NPB_554,2000_Kitanine_NPB_567,2000_Goehmann_JPA_33,2000_Maillet_NPB_575,2002_Kitanine_NPB_641,2009_Kitanine_JSTAT_P04003}) is known. 

The norm of an eigenstate can similarly be computed algebraically using the commutation relations between monodromy matrix entries, or as the appropriate limit of Slavnov's theorem, yielding the celebrated Gaudin formula \cite{GaudinBOOK,1981_Gaudin_PRD_23,1982_Korepin_CMP_86}.

\section{The Quench Action formalism}
\label{sec:QA}
Having introduced some of the concepts and notations we shall make use of, we will in this section offer a general description of the formalism of the Quench Action, outlining the overall logic and describing the generic conditions for its applicability.
\subsection{The Quench Action and its saddle point}
Let us go back to our original problem and consider an arbitrary wavefunction at $t=0$. We assume that from this instant onwards, the time evolution occurs according to the Bethe Ansatz-solvable Hamiltonian $H$ whose normalized eigenstates are labeled by sets of quantum numbers $\{ I \}$. The initial state is exactly decomposed in the basis of these eigenstates according to
\begin{equation}
|\Psi (t=0) \rangle = \sum_{\{ I \}} e^{-S^\Psi_{\{ I \}}}  | \{ I \} \rangle
\end{equation}
where we have defined the overlap coefficients
\begin{equation}
S^\Psi_{\{ I \}} = -\ln \langle \{ I \} | \Psi (t=0) \rangle ~\in ~\mathbb{C}.
\end{equation}
One crucial property is that since we are working with normalized states, the real parts of the overlap coefficients are bounded from below, 
\begin{equation}
\exists ~S_{min} \in \mathbb{R} \geq 0 ~| ~\Re S^\Psi_{\{ I \}} \geq S_{min} ~\forall \{ I \}
\end{equation}
and tend to infinity for states with vanishing overlap.

In the eigenbasis, the time-dependent Schr\"odinger equation is now trivially solved by using $H |\{ I \} \rangle = \omega_{\{ I \}} |\{ I \} \rangle$, allowing us to write the exact time-dependent wavefunction as
\begin{equation}
|\Psi (t) \rangle = \sum_{\{ I \}} e^{-S^\Psi_{\{ I \}} -i \omega_{\{ I \}}t }  | \{ I \} \rangle.
\end{equation}
The expectation value (\ref{eq:Obart}) which we are centrally interested in can then of course be written, without any approximation, as a double Hilbert space-sized summation
\begin{equation}
\bar{\cal O} (t) = \frac{ \sum_{\{ I^l \} } \sum_{\{ I^r\}} e^{-(S^\Psi_{\{ I^l\}})^* - S^\Psi_{\{ I^r\}} + i (\omega_{\{ I^l\}} - \omega_{\{ I^r\}})t} \langle \{ I^l \}| {\cal O} | \{ I^r \} \rangle}{\sum_{\{ I \}} e^{-2\Re S^\Psi_{\{ I \}}}}.
\label{eq:tdepO1}
\end{equation}
Such a double summation is in general too difficult to perform due to the exceedingly large number of terms it contains. Our aim here is to show that in the thermodynamic limit, under mild assumptions, this summation can be drastically simplified.

Let us begin by looking at the denominator of (\ref{eq:Obart}) or (\ref{eq:tdepO1}), namely the (time-independent) normalization of the initial state:
\begin{equation}
\langle \Psi (t) | \Psi (t) \rangle = \sum_{\{ I \} } e^{-2 \Re S^\Psi_{\{ I \}}}.
\end{equation}
Using our resolution of the identity (\ref{eq:ib1}), we can write this (using a self-explanatory notation) as
\begin{equation}
\langle \Psi (t) | \Psi (t) \rangle = \sum_{\{ \rho_i \}} \sum_{\{ c_i\}} e^{-2 \Re S^\Psi_{\{\rho_i\}; \{c_i\}}}.
\label{eq:normsum2}
\end{equation}
If we are using normalized states at all steps, this summation is of course equal to one. The detailed distribution of contributions is however not trivial, and the weight distribution is utterly undemocratic. For a generic initial state, the overlaps can be rapidly-varying numbers as we move around the Hilbert space. Modifying the quantum number of a single particle even by the smallest allowable unit can for example change the overlap by a factor of order one. In fact, the overlaps can even suddenly identically vanish if some discrete symmetry requirement is violated. Assuming all such discrete symmetries have been handled by a proper partitioning of the Hilbert space\footnote{A concrete example being the parity-invariance requirement in the BEC to Lieb-Liniger quench \cite{2014_DeNardis_PRA_89}, which we will discuss in more detail later on.}, the remaining overlaps do not by any means have to be smooth functions of the state's quantum numbers. On the other hand, one can expect that the extensive part of the logarithm of the overlaps {\it is} insensitive to the microscopic details of the state, and rather depends only on the overall root distribution. To express this formally, for a given set of box fillings $\{ \rho_i \}$, we can define an effective box-averaged overlap according to
\begin{equation}
e^{- S^o_{\{ \rho_i\}}} \equiv e^{-S^{YY}_{\{ \rho_i\}}}\sum_{\{ c_i\}} e^{-2 \Re S^\Psi_{\{\rho_i\}; \{c_i\}}} 
\end{equation}
in which $S^o_{\{ \rho_i\}}$ is a real-valued function of the fillings $\{ \rho_i \}$, representing the overlaps, which becomes
a well-defined, differentiable functional of the smooth distribution $\rho_{sm}$ (\ref{eq:rhosm}) in the thermodynamic limit, 
\begin{equation}
\lim_{Th,reg} S^o_{\{ \rho_i \}} \equiv S^o [\rho_{sm}].
\label{eq:So}
\end{equation}
The same smoothness in the thermodynamic limit is of course automatically true (and traditionally tacitly assumed) of the Yang-Yang entropy,
\begin{equation}
\lim_{Th,reg} S^{YY}_{\{ \rho_i \}} \equiv S^{YY} [\rho_{sm}].
\end{equation}

Once the in-box summations have been performed in this way, the remaining summation over box fillings can be interpreted as a conventional functional integral over continuously differentiable functions $\rho_{sm}$, 
\begin{equation}
\lim_{Th,reg} \sum_{\{ \rho_i \}} (...) = \int_{\rho_{sm} \in C^\infty} D\rho_{sm} (...)
\end{equation}
(since we are still working with distributions in quantum number space, where statistical independence holds, there is no extra Jacobian in this functional integral) so the normalization summation (\ref{eq:normsum2}) can be rewritten as 
\begin{equation}
\lim_{Th,reg} \langle \Psi (t) | \Psi (t) \rangle = \int D\rho_{sm} ~ e^{-S^Q[\rho_{sm}]}
\label{eq:denomfi1}
\end{equation}
where the `Quench Action' functional is defined as the difference between the pseudo-energy obtained from the overlaps and the Yang-Yang entropy of the state,
\begin{equation}
S^{Q}[\rho] = S^o[\rho] - S^{YY}[\rho].
\label{eq:QA}
\end{equation}
By construction, the Quench Action thus represents a sort of equivalent of a free energy for out-of-equilibrium situations. It is an extensive (any non-extensive part takes on large positive values, and thus does not influence the expectation values considered), real-valued functional which is bounded from below due to the state normalization constraint. One of the appeals of the Quench Action approach is that it is amenable to all the standard field-theoretical methods we are accustomed to in (quantum) statistical mechanics. This will be discussed in more detail later on in the perspectives.

As a first step towards a more useful representation, let us apply a steepest-descent reasoning in the evaluation of (\ref{eq:denomfi1}). Assuming that there exists a single minimum\footnote{The generalization to many distinct minima is straightforward; that to the case of degenerate manifolds slightly less so, although one can then follow inspirations from traditional statistical mechanics.}, the functional integral (\ref{eq:denomfi1}) can be evaluated in a saddle-point approximation (the applicability of the saddle-point logic resting on the system size going to infinity),
\begin{equation}
\int D\rho_{sm} ~ e^{-S^Q[\rho_{sm}]} = e^{-S^Q[\rho_{sp}]} ~\mbox{Det} \left[ \frac{T}{2\pi} \right]^{-1/2} \left(1 + C_L \right)
\end{equation}
where $C_L$ represent corrections vanishing in the thermodynamic limit, and $T(\lambda, \mu) = \left. \frac{\delta^2 S^Q[\rho]}{\delta \rho(\lambda) \delta \rho(\mu)} \right|_{\rho_{sp}}$ is the functional Hessian of the Quench Action evaluated at the (continuously differentiable) saddle-point distribution $\rho_{sp}$, which is determined as the distribution that satisfies the generalized TBA (GTBA, see for example \cite{2012_Caux_PRL_109,2012_Mossel_JPA_45}) equilibrium condition
\begin{equation}
\left. \frac{\delta S^Q[\rho]}{\delta \rho} \right|_{\rho_{sp}} = 0.
\label{eq:rhosp}
\end{equation}
The Hessian (provided it doesn't vanish) gives only subleading contributions in the thermodynamic limit, and can be neglected if one focuses on the dominant contributions only. Including subdominant terms is procedurally straightforward from the construction above.

In practice, the equations derived from the saddle-point condition (\ref{eq:rhosp}) are morphologically identical to the thermodynamic Bethe Ansatz equations one obtains when treating finite-temperature equilibrium integrable models. In TBA (see \cite{1969_Yang_JMP_10}, \cite{TakahashiBOOK} and references therein), one indeed performs a saddle-point analysis, which becomes exact in the thermodynamic limit, the functional integral weight being simply the (exponential of minus the) free energy of the system. The Quench Action (\ref{eq:QA}) however contains terms which depend on the density function $\rho$ in ways distinct from the thermal free energy, meaning that the `driving terms' in equations (\ref{eq:rhosp}) are distinct from the usual TBA ones. Explicit examples of these will be given in the next section.

Now is a good time to discuss a subtle point: is the Quench Action (\ref{eq:QA}) equivalent to the GGE? The Quench Action is by construction a mathematically exact representation of the diagonal ensemble; the GGE, when implemented correctly, converges to it in the thermodynamic limit. If the saddle-point approximation can be used in both cases to find a steady state, then these steady states must be the same. In this sense, equivalence is obvious. That said, the fact that the QA and the GGE share the saddle point does not mean that beyond-saddle-point features must coincide. More importantly, the point is that the QA and the GGE are founded on different footholds, meaning that approximations within one approach will not be expressible within the language of the other. For example, approximating the GGE by truncating the set of charges used has no meaningful equivalent within the QA approach. Conversely, a clever approximation scheme for evaluating the extensive parts of the overlap logarithms has no obvious translation to the language of conserved charges. Thus, though properly-implemented and performed calculations within the QA and GGE schemes should agree on the steady state, the two approaches remain quite distinct on the practical level. A more fundamental difference between the approaches is however discussed in the next subsection.

\subsection{Time-dependent expectation values}
Let us now tackle the more challenging (but physically rich and interesting) question of the time evolution.
In the context of integrable models, much work has already been performed on this issue, see for example
\cite{2006_Cazalilla_PRL_97,2009_Iucci_PRA_80,2010_Iucci_NJP_12,2010_Mossel_NJP_12,2011_Calabrese_PRL_106,2012_Calabrese_JSTAT_P07016,2012_Calabrese_JSTAT_P07022,2012_Mossel_NJP_14,2012_Schuricht_JSTAT_P04017,2012_Iyer_PRL_109,2013_Iyer_PRA_87,2013_Mussardo_PRL_111,2013_Collura_PRL_110,2013_Collura_JSTAT_P09025,2014_Delfino_JPA_47,2014_Kormos_PRA_89,2014_Rajabpour_PRA_89,2015_Rajabpour_PRB_91,2015_Kormos_1507.02708,2016_Pozsgay_1602.03065}.

Within the Quench Action formalism, obtaining the time dependence involves evaluation of the numerator of (\ref{eq:tdepO1}), 
\begin{equation}
\langle \Psi (t) | {\cal O} | \Psi (t) \rangle = \sum_{\{ I^l \}} \sum_{\{ I^r \}} e^{-(S^\Psi_{\{ I^l \}})^* - S^\Psi_{\{ I^r\} }} e^{i (\omega_{\{ I^l \}} - \omega_{\{ I^r \}}) t} \langle \{ I^l \} | {\cal O} | \{ I^r \} \rangle.
\end{equation}
In the thermodynamic limit, we expect that only a minority of states significantly contribute to this double sum. A major simplification comes from exploiting the fact that for a typical physical operator ${\cal O}$ (local density, particle addition/removal, etc.), the matrix elements $\langle \{ I^l \} | {\cal O} | \{ I^r \} \rangle$ are rapidly-decreasing functions of the `difference' between the bra and ket states, in other words of the number of displaced quantum numbers from one state to the other. We will call the operator ${\cal O}$ a `weak' operator if its matrix elements
$\langle \{ I^l \} | {\cal O} | \{ I^r \} \rangle$ are negligible unless 
$
\{ I^l \} = \{ I^r \} + \mbox{excitations carrying zero entropy}.
$
In other words, a weak operator ${\cal O}$ does not produce finite-entropy modifications of the state it acts upon.
An equivalent way of saying this is that the linear decomposition 
\begin{equation}
{\cal O} | \{ I \} \rangle = \sum_{{\bf e} } o_{\{ I \}; {\bf e}} | \{ I \}; {\bf e} \rangle, \hspace{10mm} o_{\{ I \}; {\bf e}} \in \mathbb{C}
\end{equation}
representing the action of ${\cal O}$ on a certain state can be truncated to a discrete, sub-entropically large number of excitations as one approaches the thermodynamic limit, while obeying all available sum rules to arbitrary accuracy. 
To put it quantitatively, let us use the operator ${\cal O}$'s normalized elements 
\begin{equation}
n_{\{ I \}; {\bf e} } \equiv o_{\{ I \}; {\bf e} } \left[\sum_{{\bf e} } |o_{\{ I \}; {\bf e} }|^2\right]^{-1/2}
\end{equation}
to define a Shannon-like operator entropy
\begin{equation}
S^{\cal O}_{\{ I \}} \equiv -  \sum_{{\bf e}} \left( |n_{\{ I \}; {\bf e} }|^2 \ln |n_{\{ I \}; {\bf e} }|^2 + (1 - |n_{\{ I \}; {\bf e} }|^2 ) \ln (1 - |n_{\{ I \}; {\bf e} }|^2 ) \right).
\end{equation}
A weak operator is then an operator which has a subextensive operator entropy. Note that this statement depends on both the operator and on the state on which it is applied, and is moreover basis-dependent: the operator entropy is calculated in the eigenstates basis (since we are interested in dephasings under time evolution). It implicitly depends on the correspondence between the structure of the constituents of ${\cal O}$ (for example, ${\cal O}$ could be a simple sum of local operators) and the operators corresponding to the creation of eigenstates (namely, products of $B$ operators). If these are similar, the operator entropy is low. One also generally expects a finite product of weak operators to itself be weak.
Note that multiple-point, time-split operator insertions like $e^{i (t_a - t_b) H} {\cal O}_a e^{-i(t_a - t_b)H}{\cal O}_b$ representing dynamical correlations, are also weak operators if ${\cal O}_a$ and ${\cal O}_b$ are weak. 

Assuming that ${\cal O}$ is a weak operator, the numerator of (\ref{eq:tdepO1}) simplifies to
\begin{eqnarray}
\langle \Psi (t) | {\cal O} | \Psi (t) \rangle = \sum_{\{ \rho_i \} } \sum_{\{ c_i \}} \sum_{{\bf e}} e^{-(S^\Psi_{ \{ \rho_i \}; \{ c_i\}})^* -S^\Psi_{ \{ \rho_i \}; \{ c_i\}; {\bf e}}} \times \nonumber \\
\times e^{i (\omega_{ \{ \rho_i \}; \{ c_i\}} - \omega_{\{ \rho_i \}; \{ c_i\}; {\bf e}}) t} \langle \{ \rho_i \}; \{ c_i\} | {\cal O} | \{ \rho_i \}; \{ c_i \}; {\bf e} \rangle.
\end{eqnarray}
To proceed further, we note that the thermodynamically finite energy difference between the bra and ket states depends only on the discrete excitations and the smooth density $\rho_{sm}$, but not on the in-box configurations $\{ c_i\}$. This is a very general and robust feature of states in the thermodynamic limit. We will write
\begin{eqnarray}
\lim_{Th,reg}(\omega_{ \{ \rho_i \}; \{ c_i\}; {\bf e}} - \omega_{\{ \rho_i \}; \{ c_i\}})
\equiv \omega_{{\bf e}}[\rho_{sm}].
\end{eqnarray}
Since the energy- and time-dependent phase can now be factorized out of the $\{ c_i \}$ configuration sums, we can define box-averaged combinations of the overlaps and matrix element products. To do this, we assume that the weak operator ${\cal O}$ we consider is such that its matrix elements are invariant under a simultaneous shift of in-box configurations in the bra and ket states, up to corrections which vanish in the thermodynamic limit. 
We will then call the operator `smooth'. Symbolically, we can write that a smooth operator ${\cal O}$ is such that
\begin{equation}
\lim_{Th,reg} \frac{\langle \{ \rho_i \}; \{ c_i'\} | {\cal O} | \{ \rho_i \}; \{ c_i' \}; {\bf e} \rangle}{
\langle \{ \rho_i \}; \{ c_i\} | {\cal O} | \{ \rho_i \}; \{ c_i \}; {\bf e} \rangle} = 1 + o(1)
\hspace{5mm} \forall ~\{ c_i\}, \{ c_i' \}.
\end{equation}
Equivalently, a smooth operator ${\cal O}$ commutes with operators enforcing only in-box modifications of the quantum numbers, up to terms vanishing in the regularized thermodynamic limit. If we define in-box shuffling operators
\begin{equation}
P_{\{ c_i\} \{ c_i'\} } \equiv | \{ c_i\} \rangle \langle \{ c_i'\} |,
\end{equation}
the statement that operator ${\cal O}$ is smooth is equivalent to the vanishing of the commutation of ${\cal O}$ with all in-box shufflings, as far as all expectation values are concerned, 
\begin{equation}
\langle \{ \rho_i \} | \left[ {\cal O}, P_{\{ c_i\} \{ c_i'\} } \right] | \{ \rho_i \} \rangle = o(1) \hspace{10mm} \forall ~\{ c_i\}, \{ c_i'\}.
\end{equation}

The assumption that ${\cal O}$ is a weak, smooth operator allows us to considerably simplify our equations by picking a particular in-box configuration as a representative and to take the matrix element out of the in-box configuration sum, writing
\begin{eqnarray}
\sum_{\{ c_i \}} e^{-(S^\Psi_{ \{ \rho_i \}; \{ c_i\}})^* -S^\Psi_{ \{ \rho_i \}; \{ c_i\}; {\bf e}}}
\langle \{ \rho_i \}; \{ c_i\} | {\cal O} | \{ \rho_i \}; \{ c_i \}; {\bf e} \rangle
\nonumber \\
\equiv e^{-S^o_{\{ \rho_i\} } + S^{YY}_{\{ \rho_i \}}} e^{-\delta S_{\{\rho_i\}; {\bf e}}} \langle \{ \rho_i \}; \{ 0 \} | {\cal O} | \{ \rho_i \}; \{ 0 \}; {\bf e} \rangle
\end{eqnarray}
where the box average of the overlap products
\begin{equation}
e^{-\delta S_{\{ \rho_i\}; {\bf e}}} \equiv e^{S^o_{\{ \rho_i\}} -S^{YY}_{\{ \rho_i \}}} \sum_{\{ c_i\}} e^{-(S^\Psi_{ \{ \rho_i \}; \{ c_i\}})^* - S^\Psi_{ \{ \rho_i \}; \{ c_i\}; {\bf e}}}
\end{equation}
is well-defined and naturally becomes a smooth functional of $\rho_{sm}$ and function of the discrete excitations in the regularized thermodynamic limit, 
\begin{equation}
\lim_{Th,reg} \delta S_{\{ \rho_i\}; {\bf e}} \equiv \delta S_{{\bf e}} [\rho_{sm}]
\label{eq:dSexc}
\end{equation}
with condition $\delta S_{0}[\rho] = 0$ by definition.
Using this `differential overlap' $\delta S$, the thermodynamic limit of our expression for the time-dependent correlator thus becomes, up to vanishing corrections,
\begin{eqnarray}
\fl
\lim_{Th,reg} \bar{\cal O} (t) = 
\frac{
\int {\cal D} \rho_{sm} e^{-S^{Q}[\rho_{sm}]} ~\lim_{Th,reg} \sum_{{\bf e}} e^{-\delta S_{{\bf e}}[\rho_{sm}] - i \omega_{{\bf e}}[\rho_{sm}] t} \langle \rho_{sm} | {\cal O} | \rho_{sm}; {\bf e} \rangle
}
{\int {\cal D} \rho_{sm} ~e^{-S^{Q}[\rho_{sm}]} }
\label{eq:tdepOa}
\end{eqnarray}
where we interpret the summation in the numerator as the thermodynamic limit of a summation regularized by performing it at a large but not infinite size (in order to keep the matrix elements individually finite). 
$\langle \rho_{sm} | {\cal O} | \rho_{sm}; {\bf e} \rangle$ 
is thus viewed at this stage as the matrix element obtained in a arbitrary but fixed regularization of $\rho_{sm}$ in terms of a box-regularized $\{ \rho_i \}$ and a fixed choice of in-box configurations $\{ c_i\}$, for example 
$\langle \{ \rho_i \}; \{ 0 \} | {\cal O} | \{ \rho_i \}; \{ 0 \}; {\bf e} \rangle$. 
These matrix elements are completely defined (exactly, irrespective of the interaction parameter, to all orders in inverse system size) by their algebraic Bethe Ansatz representation \cite{1990_Slavnov_TMP_82,1997_Kojima_CMP_188,1999_Kitanine_NPB_554,2000_Kitanine_NPB_567,2007_Caux_JSTAT_P01008,2007_Castro_Alvaredo_JPA_40,2011_Pozsgay_JSTAT_P01011,2012_Pozsgay_JPA_45,2013_Belliard_JSTAT_P04033,2014_Pakuliak_NPB_881,2015_Piroli_JPA_48,2015_Pakuliak_JPA_48} (typically in terms of a determinant or summations thereof), and thus completely free of singularities. They are however very strongly dependent on the relative microscopic positioning of the quantum numbers between the bra and ket states, and thus cannot be viewed as a `smooth', slowly varying function of the relevant parameters here. The summation over ${\bf e}$ should be interpreted at this stage as running over all excitations defined by allowable changes in the quantum number configuration for a given microscopic regularization, which essentially splits up into in-box entropy-like summations and dispersing excitation summation. This last summation itself has however a completely well-defined thermodynamic limit. We discuss this point in more detail later on in subsection \ref{subsec:FT}.

Formula (\ref{eq:tdepOa}) is already considerably simplified as compared to the original starting point (\ref{eq:Obart}), in the sense that only a single functional integral remains. It is of quite general applicability, being valid for arbitrary times $t$ and any smooth operator ${\cal O}$. Under mild assumptions, we can however proceed further using a saddle-point evaluation. We can assume that the differential overlap
$\delta S_{{\bf e}}[\rho_{sm}]$, 
encoding the correlation between quench overlaps of nearby states in Hilbert space, is non-extensive and thus does not (to leading order in system size) shift the saddle-point of the numerator as compared to that of the denominator. Note that this assumption is completely natural: modifying the quantum numbers of a single particle typically modifies the overlap by a factor which is algebraic rather than exponential in system size. Then, for operators ${\cal O}$ whose matrix elements are not exponentially large in system size (and thus also do not shift the saddle point; we will then call ${\cal O}$ {\it thermodynamically finite}), we can use the previously-obtained saddle point $\rho_{sp}$ (\ref{eq:rhosp}) and get the yet further significantly simplified form (explicitly reintroducing for convenience the symmetry between putting the excitations in the bra or the ket, which is implicitly present but not manifest in (\ref{eq:tdepOa}))
\begin{eqnarray}
\fl
\lim_{Th,reg} \bar{\cal O} (t) = \lim_{Th,reg} \frac{1}{2} \sum_{{\bf e}} \left[ e^{-\delta S_{{\bf e}}[\rho_{sp}] - i \omega_{{\bf e}}[\rho_{sp}] t} \langle \rho_{sp} | {\cal O} | \rho_{sp}; {\bf e} \rangle \right. \nonumber \\
\hspace{3cm}\left. + e^{-\delta S^*_{{\bf e}}[\rho_{sp}] + i \omega_{{\bf e}}[\rho_{sp}] t} \langle \rho_{sp}; {\bf e} | {\cal O} | \rho_{sp} \rangle \right].
\label{eq:tdepOsp}
\end{eqnarray}
This equation is perhaps the most crucial formula of the Quench Action formalism. It is applicable provided operator ${\cal O}$ is
\begin{enumerate}
\item weak,
\item smooth,
\item thermodynamically finite
\end{enumerate}
(in particular, note that these conditions are fulfilled quite generally by local operators).
Note the crucial fact that we have not assumed large times while deriving this formula; in fact, we conjecture that expression (\ref{eq:tdepOsp}), evaluated in the thermodynamic limit, is valid for {\it all} times $t > 0$ provided the operator ${\cal O}$ is weak, smooth and thermodynamically finite. 
At large times, meaning at times much larger than the lowest considered excitation energy (that being of order $1/L$, and not $e^{-(cst)L}$ like the mean energy spacing), the formula simplifies to its diagonal part
\begin{equation}
\lim_{Th,reg} \bar{\cal O} (t) = \lim_{Th,reg} \langle \rho_{sp} | {\cal O} | \rho_{sp} \rangle
\label{eq:tdepOsplarget}
\end{equation}
with right-hand side evaluated on any microscopic realization of the saddle-point state. 
This parallels the microcanonical sum used in \cite{2011_Cassidy_PRL_106}, which can be interpreted as a generalization of the Eigenstate Thermalization Hypothesis \cite{1991_Deutsch_PRA_43,1994_Srednicki_PRE_50,1996_Srednicki_JPA_29,1999_Srednicki_JPA_32,2008_Rigol_NATURE_452,2009_Rigol_PRL_103}. 

In terms of computational complexity, the representation (\ref{eq:tdepOsp}) thus displays the fact that the Quench Action `starts from the steady state' and works its way backwards in time upon the addition of more and more excitations around the steady state.

\subsection{Simplifications from additivity}
Yet a further slightly simplified version of (\ref{eq:tdepOsp}) can be obtained by exploiting 
the fact that energies of the individual discrete excitations around the saddle point are additive. This is a well-known general feature of the thermodynamic limit of the exact eigenstates we are playing with. More precisely, 
if ${\bf e}$ represents 
the set of excitations (particles, holes, etc) 
which have been created on the state described by the thermodynamic root distribution $\rho_{sp}$, we have
\begin{equation}
\omega_{{\bf e}}[\rho_{sp}] = \sum_i \varepsilon (e_i) 
\end{equation}
in which 
$\varepsilon(e_i)$ 
is a single function representing the dispersion relation for excitations around the saddle point (and obtainable via the GTBA). 

We can even go one step further and assume that, similarly to the energies, the differential overlap splits into decoupled functions of the individual excitations\footnote{Formally, the notions here can be shown to converge for the box-averaged states introduced in subsection \ref{subsec:FT}.}, namely that there exists a single complex-valued function $s$ such that
\begin{equation}
\delta S_{{\bf e}} [\rho] = \sum_i s (e_i) + ...
\label{eq:overlapfn1}
\end{equation}
encoding the overlap differences between states appearing in the sum (\ref{eq:tdepOsp}).
Such a function can be read from considering the scaling of the exact overlaps as system size goes to infinity, an example being formula (\ref{eq:BECdiffoverlaps}) (from (A14) of \cite{2014_DeNardis_PRA_89}). In fact, one cannot a priori exclude the possibility that (\ref{eq:overlapfn1}) can be oversimplified, and that one should include a whole series of additional many-body terms 
\begin{equation}
\delta S_{{\bf e}} [\rho] = \sum_i s(e_i) + \sum_{i_1 < i_2} s^{(2)}(e_{i_1}, e_{i_2}) + ...
\end{equation}
In most situations however the two-body term (and higher ones) would carry only finite-size corrections. 
The fact remains that the function $s$ (and eventually it's many-body corrections) can be obtained directly from the exact overlaps. We will provide examples of this in the next section.

We can in fact push the preceding logic one step further. Let us examine in some more detail the required matrix elements
\begin{equation}
\langle \rho_{sp} | {\cal O} | \rho_{sp}; {\bf e} \rangle.
\label{eq:ME}
\end{equation}
Physically relevant operators are such that the value of such a matrix element decreases upon inserting more and more excitations. Moreover, the factor by which the matrix element decreases upon adding an excitation, depends primarily on the added excitation, with additional corrections coming from the interplay between excitations. This leads us to propose to represent the matrix elements (\ref{eq:ME}) in terms of weights $h$ as
\begin{equation}
\langle \rho_{sp} | {\cal O} | \rho_{sp}; {\bf e} \rangle = e^{- h^{\cal O}_{\rho_{sp}; {\bf e}}},
\end{equation}
these weights taking complex values in general; for a normalized operator, the real part is however naturally bounded from below. 
One can of course `lift' this representations to the operator level, writing ${\cal O} = e^{-\hat{h}^{\cal O}_{\rho_{sp}}}$ in terms of the `operator pseudo-Hamiltonian' $\hat{h}^{\cal O}_{\rho_{sp}}$ (note that this pseudo-Hamiltonian depends on both the operator and saddle-point state). Somewhat in parallel to the Quench Action itself, the so-defined operator pseudo-Hamiltonian is a measure of the importance of eigenstates as far as expectation values are concerned. Important states have small (real part of) pseudo-energy; conversely, states with negligible matrix elements correspond to high pseudo-energy states.
The potentially crucial simplification comes if the weights can then be expressed (as done above for the differential overlaps) in terms of one-, two-, ... body functions,
\begin{equation}
h_{{\bf e}} = \sum_i h(e_i) + \sum_{i_1 < i_2} h^{(2)}(e_{i_1}, e_{i_2}) + ...
\end{equation}
An operator which is not too dissimilar to the operator creating individual particle-hole excitations would naturally be represented using only the first few terms to achieve good accuracy. The advantage of looking for such a representation is that the whole operator structure is then encoded in the smallest number of parameters possible. This topic, which has not yet been implemented beyond unpublished isolated cases and goes beyond the Quench Action towards the computation of correlations on generic states, will be investigated further in future works.

\vspace{3mm}
To summarize, in order to reconstruct the full time dependence of the expectation value of the weak, smooth operator ${\cal O}$, the only ingredients that are needed are:
\begin{enumerate}
\item the saddle-point distribution $\rho_{sp}$, 
\item the excitation energy function $\varepsilon$\footnote{Note that this is equivalent to specifying the Hamiltonian; the saddle-point distribution $\rho_{sp}$ specifies the Bethe state completely, and thus also the structure of the excitations in its vicinity, including their density of states. Given a Hamiltonian and a $\rho_{sp}$, all $\varepsilon$ can be computed. Conversely, given a $\rho_{sp}$ and $\varepsilon$, all energies are known (and thus so is the Hamiltonian) for states `in the vicinity' of the saddle-point.}
\item the characteristic quench overlap function $s$ (and if needed its higher-body parts $s^{(2)}$ etc.), and
\item the matrix elements $\langle \rho_{sp} | {\cal O} | \rho_{sp}; \{ {\bf e} \} \rangle$ for states around the saddle point, either directly or in terms of the relevant operator's pseudo-Hamiltonian.
\end{enumerate}
These can be viewed as the distilled `effective parameters' encoding the whole time evolution of a weak, smooth operator after the quench. Note that the first three items are operator-independent.

\subsection{Further steps towards a more field theory-like language}
\label{subsec:FT}
It is worthwhile here to discuss how one should properly take the thermodynamic limit of all our constructions up to now. The point to remember is that the underlying integrable model provides a fully regularized theory at all energies, to all orders in system size, for all states and operators. The finite $N$ equations thus hide all the necessary details of a regularization scheme which needs to be appended to the field theory in order to make it well-defined. Though we discuss this aspect here in the context of the Quench Action and its application, the same reasonings are applicable in much more generality.

As explained above, the actual calculation of (\ref{eq:tdepOsp}) should be performed using a defined microscopic realization of the steady state, for example the `maximally flat' one,
{ \setlength{\mathindent}{5mm}
\begin{equation}
\sum_{\bf e} e^{-\delta S_{{\bf e}}[\rho_{sp}] - i \omega_{{\bf e}}[\rho_{sp}] t} \langle \rho_{sm} | {\cal O} | \rho_{sm}; {\bf e} \rangle 
= \sum_{\bf e} e^{-\delta S_{{\bf e}}[\rho_{sp}] - i \omega_{{\bf e}}[\rho_{sp}] t} \langle \{ \rho_i \}; \{ 0 \} | {\cal O} | \{ \rho_i \}; \{ 0 \}; {\bf e} \rangle.
\label{eq:FTsum1}
\end{equation}
}\noindent
Let us specify the structure of the sum over excitations a little bit more precisely. In our box regularization, let us as before use latin indices $i$ to label the boxes. The difference between the bra and ket states in (\ref{eq:FTsum1}) consists in in-box `entropy-like' modifications, accompanied by out-box `dispersing' excitations taking the form of particles or holes drilled on the overall density profile $\{ \rho_i \}$. It is convenient to separate these two types of excitations explicitly, and to effectively perform a summation over the former. Denoting the number of such dispersing particle (resp. hole) excitations as $n_p$ (resp. $n_h$) and the box which they disperse to by $i_a, a = 1, ..., n_p$ (resp. $\bar{i}_b, b=1, ..., n_h$), we can represent the sum over excitations in the example above as
{ \setlength{\mathindent}{10mm}
\begin{eqnarray}
\sum_{\bf e} e^{-\delta S_{{\bf e}}[\rho_{sp}] - i \omega_{{\bf e}}[\rho_{sp}] t} \langle \{ \rho_i \}; \{ 0 \} | {\cal O} | \{ \rho_i \}; \{ 0 \}; {\bf e} \rangle = \nonumber \\
\sum_{n_p = 0}^\infty \sum_{n_h = 0}^\infty \frac{1}{n_p! n_h!} 
\sum_{\{ i_a \}_{n_p} ; \{ \bar{i}_b \}_{n_h}}
\left[\prod_{a=1}^{n_p} \prod_{b=1}^{n_h} (1 - \delta_{i_a \bar{i}_b})\right] 
e^{- \sum_a s_{i_a} + \sum_{b} s_{\bar{i}_b} -i (\sum_a \varepsilon_{i_a} - \sum_{b} \varepsilon_{\bar{i}_b}) t}
\times \nonumber \\
\times 
\sum_{\{ c_i \}} \langle \{ \rho_i \}; \{ 0 \} | {\cal O} | \{ \rho_i \} \cup \{ i_a \}, \{ \bar{i}_b \}; \{ c_i \} \rangle.
\end{eqnarray}
}\noindent
A few things are worth clarifying. First of all, $\sum_{\{ c_i \}}$ represents as before the in-box summations at fixed density profile $\{ \rho_i \}$. The factors $1 - \delta_{i_a \bar{i}_b}$ enforce the convention that dispersing particle and hole excitations are chosen by convention not to land in the same box, this being meaningless since the density set $\{ \rho_i \}$ is then not modified (the sum over configurations $\{ c_i \}$ already takes care of these terms). Multiple particles (resp. holes) can however land in the same box. We have also exploited the additivity of the differential overlap and energy functions around the saddle point. 

It is convenient at this stage to introduce `box-averaged' states
\begin{equation}
| \{ \rho_i \} \rangle_{b} \equiv e^{-\frac{1}{2} S^{YY}_{\{ \rho_i\}} } \sum_{\{ c_i \}} | \{ \rho_i \}; \{ c_i \} \rangle
\label{eq:boxavgstates}
\end{equation}
obtained by uniformly summing over all in-box configurations at fixed density. Box-averaged states are not exact eigenstates of our Hamiltonian; nevertherless, their energy is well-defined with fluctuations which are exponentially small in system size and can be neglected in the thermodynamic limit. Box-averaged states are still orthonormal in the sense that
\begin{equation}
{}_{b}\langle \{ \rho^l_i \} | \{ \rho^r_i \} \rangle_{b} = \delta_{\{ \rho^l_i\}, \{\rho^r_i\}}.
\end{equation}
One can also introduce for convenience the usual ladder operators $z_j, z^\dagger_j$ such that
\begin{equation}
z^\dagger_j | \{ \rho_i \} \rangle_b = | \{ \rho_i + \frac{\delta_{ij}}{l_j} \} \rangle_b, \hspace{10mm}
z_j | \{ \rho_i \} \rangle_b = | \{ \rho_i - \frac{\delta_{ij}}{l_j} \} \rangle_b
\end{equation}
with simple algebra and additional lowest/highest-weight conditions
\begin{equation}
\left[ z_j, z^\dagger_{j'} \right] = \delta_{j j'}, \hspace{10mm} z^\dagger_j | \{ \rho_i \} \rangle_b|_{\rho_j = l_j} = 0, \hspace{10mm}
z_j | \{ \rho_i \} \rangle_b|_{\rho_j = 0} = 0.
\end{equation}
These simple operators are somewhat reminiscent of Zamolodchikov-Faddeev operators in integrable field theory \cite{1979_Zamolodchikov_AP_120} (these operators have indeed been used in the context of quenches, see {\it e.g.} \cite{2010_Fioretto_NJP_12,2012_Sotiriadis_JSTAT_P02017}). We however want to emphasize that they are not the same: the simple operators here are still fully microscopics-aware, and permit calculations in principle to all orders in system size (to put it in field theory language: there is no need for further regularization, there are no infinities present at this stage).

Introducing operators $f_I$, $f^\dagger_I$ defined as operators removing/adding an occupation at quantum number $I$ in the full microscopic model, and the in-box symmetric combination $f_j \equiv \sum_{I \in \mbox{\tiny box} ~j} f_I$ (one can think of the (un)occupied quantum numbers as a pseudo-spin-$1/2$ degree of freedom, with operators $f_I/f^\dagger_I$ as a spin lowering/raising operator, and the $f_j, f^\dagger_j$ as total spin operators in the representation with total spin $l_j/2$), we have the actions (taking the factors coming from the entropy into account)
\begin{eqnarray}
f^\dagger_j |\{ \rho_i \} \rangle_b = \left[ (n_j + 1) (l_n - n_j) \right]^{1/2} | \{ \rho_i + \frac{\delta_{ij}}{l_j} \} \rangle_b, \nonumber \\
f_j |\{ \rho_i \} \rangle_b = \left[ n_j (l_j - n_j + 1) \right]^{1/2} | \{ \rho_i - \frac{\delta_{ij}}{l_j} \} \rangle_b
\end{eqnarray}
and thus the identities
{ \setlength{\mathindent}{5mm}
\begin{equation}
z^\dagger_j = f^\dagger_j \frac{l_j}{\left[ (\hat{\rho}_j + \frac{1}{l_j}) (1 - \hat{\rho}_j) \right]^{1/2}} ~~(\rho_j \neq 1), \hspace{5mm}
z_j = f_j \frac{l_j}{\left[ \hat{\rho}_j (1 - \hat{\rho}_j + \frac{1}{l_j}) \right]^{1/2}} ~~(\rho_j \neq 0)
\end{equation}
}\noindent
in which $\hat{\rho}_j$ returns the density in box $j$. The use of introducing box-averaged states is that these will become the well-defined states in the thermodynamic limit, parametrized by the density distribution. We can then define a box-renormalized operator ${\cal O}^{br}$, with matrix elements in the box-averaged state basis absorbing the renormalization coming from the in-box summation,
{ \setlength{\mathindent}{5mm}
\begin{eqnarray}
\frac{1}{\prod_{a=1}^{n_p} (l_{i_a} - n_{i_a}) \prod_{b=1}^{n_h} n_{\bar{i}_b}} \sum_{\{ c_i \}} \langle \{ \rho_i \}; \{ 0 \} | {\cal O} | \{ \rho_i \} \cup \{ i_a \}, \{ \bar{i}_b \}; \{ c_i \} \rangle \nonumber \\
= \frac{e^{-S^{YY}_{\{ \rho_i \}}}}{\prod_{a=1}^{n_p} (l_{i_a} - n_{i_a}) \prod_{b=1}^{n_h} n_{\bar{i}_b}} 
\sum_{\{ c_i \}, \{ c_i'\}} \langle \{ \rho_i \}; \{ c_i' \} | {\cal O} | \{ \rho_i \} \cup \{ i_a \}, \{ \bar{i}_b \}; \{ c_i \} \rangle 
\nonumber \\
\equiv {}_b \langle \{ \rho_i \} | {\cal O}^{br} | \{ \rho_i \} ; \{ i_a \}, \{ \bar{i}_b \} \rangle_b
\label{eq:FTO}
\end{eqnarray}
}\noindent
in which we have in the first step exploited the assumption that operator ${\cal O}$ is smooth. The factors $l_i - n_i$ and $n_{\bar{i}}$ are included for convenience to compensate for the fact that a particle excitation can be positioned at $l_i - n_i$ positions in box $i$, and the hole in $n_{\bar{i}}$ positions, the microscopically-defined matrix element being assumed to be insensitive to the precise positioning of the dispersing excitations at the level of resolution corresponding to the box size (this holds true in general and can be verified on a case-by-base basis).

The point is that the renormalization factor is density profile-specific, but is expected to be only weakly operator dependent and does not depend on the number or positioning of the dispersing excitations (provided there is a denumerable number of those). The summation over in-box configurations is reminiscent of the summation over soft modes used in \cite{2011_Shashi_PRB_84} (where it was done for soft modes around ground states, leading to an anomalous power-law renormalization of matrix elements; here, being performed around finite-entropy states, the renormalization can be of exponential size). Actually performing the summations representing the operator renormalization is a nontrivial task, which needs to be performed on a case-by-case basis. Note that the value of the matrix elements in this box-averaged state basis (right-hand side of (\ref{eq:FTO})) is now not bounded in the thermodynamic limit. In particular, they can develop singularities when particle and hole excitations come closer together (namely when they disperse to boxes which are closer and closer to each other). Additionally, they also display a crossing symmetry, namely an equality between putting {\it e.g.} a hole in the bra or a corresponding particle in the ket. These matrix elements thus essentially obtain properties reminiscent of some of those of form factors in integrable field theory \cite{SmirnovBOOK}. We will get back to this point in more detail in future publications.

Going back to our discussion, our summation (\ref{eq:FTsum1}) has now become
{ \setlength{\mathindent}{10mm}
\begin{eqnarray}
\sum_{n_p = 0}^\infty \sum_{n_h = 0}^\infty \frac{1}{n_p! n_h!} 
\sum_{\{ i_a \}_{n_p} ; \{ \bar{i}_b \}_{n_h}}
\left[\prod_{a=1}^{n_p} \prod_{b=1}^{n_h} (1 - \delta_{i_a \bar{i}_b})\right] 
\prod_{a=1}^{n_p} (l_{i_a} - n_{i_a}) \prod_{b=1}^{n_h} n_{\bar{i}_b} \times \nonumber \\
\times e^{- \sum_a s_{i_a} + \sum_{b} s_{\bar{i}_b} -i (\sum_a \varepsilon_{i_a} - \sum_{b} \varepsilon_{\bar{i}_b}) t}
{}_b \langle \{ \rho_i \} | {\cal O}^{br} | \{ \rho_i \} ; \{ i_a \}, \{ \bar{i}_b \} \rangle_b
\end{eqnarray}
}\noindent
in which 
\begin{equation}
| \{ \rho_i \} ; \{ i_a \}, \{ \bar{i}_b \} \rangle_b = \prod_{a} z^\dagger_{i_a} \prod_{b} z_{\bar{i}_b} | \{ \rho_i \} \rangle_b.
\end{equation}
One can then proceed with the continuum limit, introduce integrals over the quantum numbers $x = I/L$ by using $\Delta x_i = \frac{l_i}{L}$, namely $\sum_{i_a} (...) = \int dx \frac{L}{l_i} (...)$, to get (in a self-explanatory notation)
{ \setlength{\mathindent}{1mm}
\begin{eqnarray}
\sum_{n_p = 0}^\infty \sum_{n_h = 0}^\infty \frac{L^{n_p + n_h}}{n_p! n_h!} \fint \prod_{a=1}^{n_p} dx_a \prod_{b=1}^{n_h} d \bar{x}_b \prod_{a=1}^{n_p} \rho^h(x_a) \prod_{b=1}^{n_h} \rho(\bar{x}_b) 
e^{-\delta S ({\bf x}, {\bf \bar{x}}) - i \omega ({\bf x}, {\bf \bar{x}}) t} \langle \rho_{sm} | {\cal O}^{br} | \rho_{sm} ; {\bf x}, {\bf \bar{x}} \rangle \nonumber \\
\label{eq:FTdone}
\end{eqnarray}
}\noindent
in which $\fint$ means the integral with `principal part' extraction of all points $x_a = \bar{x}_b ~\forall~ a,b$. 
Note that instead of integrating over quantum number space variables, one can equivalently integrate over rapidity variables by using the transformation function $x(\lambda)$ (\ref{eq:BEext}) and its Jacobian (\ref{eq:rhosdefs}).
As described earlier, the energy $\omega$ and differential overlap function $\delta S$ are also expressible in terms of sums over excitations, the energy being purely one-body, and the differential overlap perhaps having some higher-body parts.
Note that formula (\ref{eq:FTdone}) parallels equation (2.38) in \cite{2015_DeNardis_JSTAT_P02019}, the difference resting in the precise definition of the matrix element (in our case here, box-averaged), and is an adaptation of the LeClair-Mussardo formula \cite{1999_LeClair_NPB_552} to this highly-off-vacuum context.

Expression (\ref{eq:FTdone}) and its mirror term as per (\ref{eq:tdepOsp}) then give a more traditionally field theory-looking version of the fundamental equation of the Quench Action formalism, in the sense that the excitations one sums over are strictly denumerable. Though somewhat formal, they make clear which kind of contributions need to be taken into consideration when computing the time dependence in the thermodynamic limit. They might serve as a different way to regularize quenches in integrable field theory \cite{2010_Fioretto_NJP_12,2010_Kormos_JHEP_1004,2011_Pozsgay_JSTAT_P01011,2012_Sotiriadis_JSTAT_P02017,2013_Mussardo_PRL_111,2014_Delfino_JPA_47,2014_Sotiriadis_PLB_734,2015_Essler_PRA_91,2015_Schuricht_JSTAT_P11004,2016_Cubero_JSTAT_P033115,2016_Horvath_NPB_902}.

\section{A tour d'horizon of recent applications}
\label{sec:Tour}
For the reader's orientation in the growing literature making use of the Quench Action, we here present the basic details of a number of quench situations which have been handled using its formalism. It is by no means the intention to be exhaustive or to review results in detail, but rather to give some pointers to the existing literature and a bird's eye overview of the current state of affairs.

\subsection{The transverse-field Ising model}
The first problem treated \cite{2013_Caux_PRL_110} using the general formalism of the Quench Action was that of the transverse-field Ising model
\begin{equation}
H (h) = -J \sum_{j=1}^L \left[ \sigma^x_j \sigma^x_{j+1} + h\sigma^z_j \right],
\label{eq:TFI}
\end{equation}
with $J, h > 0$. This model has two phases, one with ferromagnetic order along the $x$ direction for $h < 1$, the other being paramagnetic with $h > 1$, these being separated by a quantum critical point in the Ising universality class (see {\it e.g.} \cite{SachdevBOOK}). Bethe Ansatz is not needed here; diagonalization can be performed using Jordan-Wigner and Bogoliubov transformations, leading to free dispersive modes $\alpha_k$ with diagonal Hamiltonian 
\begin{equation}
H(h) = \sum_k \varepsilon_n (k) \left( \alpha^\dagger_k \alpha_k - \frac{1}{2} \right), \hspace{10mm}
\varepsilon_h (k) = 2J \sqrt{1 + h^2 - 2h \cos k}.
\end{equation}
The first quench protocol considered consists in preparing the system in the ground state of (\ref{eq:TFI}) for an initial value $h_0$ in the paramagnetic phase $h_0 > 1$, and to suddenly switch the field to value $h$ from $t = 0$ onwards. 

The post-quench reduced density matrix has been computed in \cite{2013_Fagotti_PRB_87}, giving $\rho^{DM} (t) = | \rho_{sp}\rangle \langle \rho_{sp}| \propto e^{\frac{1}{4}{\bf a}^T {\bf W} {\bf a}}$ with ${\bf a}$ being a vector of Majorana modes $a_{2n-1} = \left[\prod_{m<n} \sigma^z_m \right] \sigma^x_n$, $a_{2n} = \left[\prod_{m<n} \sigma^z_m \right] \sigma^y_n$ with anticommutation $\{ a_m, a_n \} = 2\delta_{mn}$, the matrix $W$ being given by $\tanh (W/2) = \Gamma$ \cite{2003_Vidal_PRL_90,2009_Peschel_JPA_42} in which
\begin{equation}
\Gamma_{jk} = ~\mbox{Tr}~\left[ \rho^{DM}(t) a_k a_j \right] - \delta_{jk}.
\end{equation}

The steady state can be obtained by applying the Quench Action method (we refer the reader to \cite{2013_Caux_PRL_110} for all details). In this case, the solution to the relevant GTBA can be obtained directly from the fact that the momentum occupations are conserved. This leads to the steady state distribution
{ \setlength{\mathindent}{10mm}
\begin{equation}
\rho(k) = \frac{1 - \cos \Delta_k}{4\pi}, \hspace{10mm} \cos \Delta_k = 4J^2 (1 + h h_0 - (h + h_0) \cos k)/[\varepsilon_h (k) \varepsilon_{h_0} (k)]
\label{eq:TFIrho}
\end{equation}
}\noindent
in which the momenta $\kappa_j$ are distributed according to the saddle-point density (\ref{eq:TFIrho}), $\kappa_{j+1} = \kappa_j + \frac{1}{L \rho(\kappa_j)}$.
This leads to a Gaussian steady state $| \rho_{sp} \rangle = \prod_j \alpha^\dagger_{\kappa_j} \alpha^\dagger_{-\kappa_j} | 0;h \rangle$ (the state $|0; h \rangle$ being the vacuum at field value $h$) with density matrix $\rho_{sp}^{DM} = |\rho_{sp} \rangle \langle \rho_{sp}|$ completely characterized by a correlation matrix whose only nonzero elements are
\begin{equation}
(\Gamma_{sp})_{2l-1, 2l-2n} = -\frac{i}{L} \sum_k \frac{e^{-ink} (h - e^{ik})(1 - 2\delta_{k, \kappa_j})}{\sqrt{1 + h^2 - 2h\cos k}}.
\end{equation}
This density matrix obeys $\Gamma_{sp} = \Gamma(t \rightarrow \infty)$ and thus indeed equation (\ref{eq:tdepOsplarget}) holds 
for any (products of) local operators, offering a simple way of reproducing the form obtained beforehand \cite{2011_Calabrese_PRL_106,2012_Calabrese_JSTAT_P07016,2012_Calabrese_JSTAT_P07022}.  For time-dependent expectation values, equation (\ref{eq:tdepOsp}) can similarly be verified at the level of the (reduced) density matrix. 

To summarize, although this quench problem had been explicitly solved before, the Quench Action logic allowed to reproduce steady-state results (and also to reobtain the known time-dependent relaxation behaviour) in a computationally less heavy fashion.

\subsection{The BEC to Lieb-Liniger quench}
The previous example pertaining to the transverse-field Ising belongs to the class of problems which, due to the absence of interactions, are solvable without invoking the technology of the Bethe Ansatz. Although the dynamics of such models is surely representative of generic situations (in particular because the mapping between interesting observables and the diagonalized modes is sometimes very complex and nonlocal) and the GGE is expected to hold, it is interesting to study situations where interactions are inevitably strong and no mapping to a simple free theory exists.
 
Let us thus now turn our attention to what is to the best of the author's knowledge the first quench problem to be solved analytically in the thermodynamic limit (by which we mean obtaining an exact analytical characterization of the steady state), for post-quench time evolution in the presence of nontrivial interactions.

In the context of bosons in one dimension, if one takes as initial state the ground state of the noninteracting bosonic gas on a periodic interval of circumference $L$, namely the Bose-Einstein condensate-like\footnote{We are abusing the term `condensate' here, since this state has no global particle number fluctuations.} state
\begin{equation}
| \Psi_0 \rangle = \frac{1}{\sqrt{L^N N!}} \left(\psi_{k=0}^\dagger\right)^N | 0 \rangle
\label{eq:BEC}
\end{equation}
in which $| 0 \rangle$ is the Fock vacuum and $\psi_{k=0}^\dagger$ creates a zero-momentum particle, one can obtain an interesting quench problem by simply turning interactions on from $t = 0$ onwards. For definiteness, the interactions are taken to be ultralocal, {\it i.e.} the post-quench Hamiltonian is defined as that of the Lieb-Liniger gas \cite{1963_Lieb_PR_130_1},
\begin{equation}
H_{LL} = - \sum_{j=1}^N \frac{\partial^2}{\partial_{x_j}^2} + 2c \sum_{j_1 < j_2} \delta (x_{j_1} - x_{j_2})
\label{eq:HLiebLin}
\end{equation}
with $c > 0$. In this case, the Bethe equations take the form
\begin{equation}
\lambda_j + \frac{1}{L} \sum_{l=1}^N \phi(\lambda_j - \lambda_l) = \frac{2\pi}{L} I_j, \hspace{10mm} j = 1, ..., N,
\end{equation}
where $\phi(\lambda) \equiv 2 ~\mbox{atan} (\lambda/c)$ and the quantum numbers are distinct half-odd integers for $N$ even and integers for $N$ odd. The wavefunction itself is given by the Bethe Ansatz,
{ \setlength{\mathindent}{10mm}
\begin{eqnarray}
\Psi_N(\{ x\} | \{ \lambda \}) &=& \prod_{N \geq j_1 > j_2 \geq 1} sgn(x_{j_1} - x_{j_2}) sgn (\lambda_{j_1} - \lambda_{j_2}) \times 
\nonumber \\
&& \times \sum_{P \in \pi_N} (-1)^{[P]} 
e^{i \sum_{j=1}^N \lambda_{P_j} x_j + \frac{i}{2} \sum_{N \geq j_1 > j_2 \geq 1} sgn(x_{j_1} - x_{j_2}) \phi (\lambda_{P_{j_1}} - \lambda_{P_{j_2}})}.
\label{eq:1DBG:BAN}
\end{eqnarray}
}\noindent

In the initial BEC state, the local number density fluctuations are large, and the momentum distribution function is by definition a delta function at zero momentum carrying the weight of all particles; turning repulsive interactions on must then lead among others to a broadening of the momentum distribution function, with the suddenly quenched interaction energy being partly converted to kinetic energy as time progresses. Since all energies are incommensurately related at finite $c$, one does not expect any persistent oscillations or recurrences in the thermodynamic limit, but rather to see the system effectively relax to a steady state due to quantum dephasing.

The search for a solution to this quench problem has an interesting history (which will be discussed later on). 
It was solved exactly in \cite{2014_DeNardis_PRA_89} using the Quench Action formalism, and what follows is a brief summary of some important aspects of this solution.

\subsubsection{Overlaps.}
In order to implement the Quench Action for this protocol, the overlaps between the BEC state (\ref{eq:BEC}) and the eigenstates (\ref{eq:1DBG:BAN}) of the Lieb-Liniger gas (\ref{eq:HLiebLin}) should be obtained. This seems, a priori, to pose insurmountable difficulties: although Bethe states are indeed composed of linear combinations of plane waves, which seem to correspond seamlessly to the eigenstates of a free Hamiltonian (and even more to the especially simple and structureless BEC state), the problem is that these plane waves are defined only on finite segments in coordinate space, and the overlap integrals always leave incommensurate phases hanging. A basic attempt at calculating the overlap using real-space integrals thus yields a factorially large sum of terms with no immediately obvious simplification in the general case. For the Lieb-Liniger gas in the Tonks-Girardeau limit, overlaps with the BEC state can be obtained \cite{2010_Gritsev_JSTAT_P05012} as products of the inverse of the rapidities.

Amazingly, for the BEC state, at any value of the interaction parameter, a dramatic simplification in fact turns out to be possible: one finds \cite{2014_DeNardis_PRA_89,2014_Brockmann_JSTAT_P05006} that the exact overlaps (which are only nonvanishing for parity-symmetric states, namely states in which the rapidity distribution is mirror-symmetric about the origin; this fact, with the Bethe equations, allows to perform simplifications) are exactly given by the relatively simple expression
\begin{equation}
\langle \Psi_0 | \{ \lambda \}_{N/2} \cup \{ -\lambda \}_{N/2} \rangle = \left[ \frac{(cL)^{-N} N!}{\mbox{det}_N G_{jk} }\right]^{1/2} \frac{\mbox{det}_{N/2} G^Q_{jk}}{\prod_{j=1}^{N/2} \frac{\lambda_j}{c} \left[ \frac{\lambda_j^2}{c^2} + \frac{1}{4} \right]^{1/2}}
\label{eq:BECoverlap}
\end{equation}
in which $G_{jk}$ is the Gaudin matrix
\begin{equation}
G_{jk} = \delta_{jk} \left[ L + \sum_{l=1}^N K (\lambda_j, \lambda_l) \right] - K(\lambda_j, \lambda_k)
\end{equation}
with kernel $K (\lambda, \lambda') = \frac{2c}{(\lambda-\lambda')^2 + c^2}$, and $G^Q$ is the Gaudin-like matrix
\begin{equation}
G^Q_{jk} = \delta_{jk} \left[ L + \sum_{l=1}^{N/2} K^Q (\lambda_j, \lambda_l) \right] - K^Q(\lambda_j, \lambda_k)
\end{equation}
with $K^Q(\lambda, \lambda') = K(\lambda, \lambda') + K(\lambda, -\lambda')$. This overlap formula is extremely econominal, in the sense that its complexity scales only with the third power of system size due to its determinant structure, instead of the factorially large worst-case expectation. It is reminiscent not only of the Gaudin norm formula, but of the matrix elements of local density or field operators for this model \cite{1989_Slavnov_TMP_79,1990_Slavnov_TMP_82,2007_Caux_JSTAT_P01008,2015_Piroli_JPA_48}.

\subsubsection{The steady state from the Quench Action.}
For the application of the Quench Action, only the extensive part of the (logarithmic) overlap is required, which can be extracted from (\ref{eq:BECoverlap}). The distribution-dependent extensive part of the overlap is
\begin{equation}
S^Q/L = \int_0^\infty d\lambda \rho(\lambda) \log\left(\frac{\lambda^2}{c^2} \left( \frac{1}{4} + \frac{ \lambda^2}{c^2}  \right) \right).
\label{eq:BECoverlapsextensive}
\end{equation}
This is then combined with the Yang-Yang entropy
\begin{equation}
S^{YY}/L =  \int_0^\infty d\lambda \left[ \rho^t(\lambda) \log\rho^t(\lambda) - \rho(\lambda) \log\rho(\lambda)  - \rho^h(\lambda) \log\rho^h(\lambda) \right]
\end{equation}
(in which $\rho^t = \rho + \rho^h$; note that this expression for the entropy reflects the required parity-invariance of the states by integrating over the positive rapidity half-line only), to give the Quench Action (\ref{eq:QA}). The variational condition (\ref{eq:rhosp}) then gives (with proper handling of the filling of the gas, see the original paper \cite{2014_DeNardis_PRA_89} for details) a functional equation for the steady-state root distribution $\rho_{sp}$. Defining $\eta(\lambda) = \frac{\rho^h(\lambda)}{\rho(\lambda)}$, this saddle-point equation reads
\begin{equation}
\ln \eta(\lambda) = g(\lambda) - h - \int_{-\infty}^\infty \frac{d\lambda'}{2\pi} K (\lambda - \lambda') \ln \left[ 1 + \eta^{-1}(\lambda') \right]
\end{equation}
and is thus of GTBA form with driving term 
\begin{equation}
g(\lambda) = \ln \left[ \frac{\lambda^2}{c^2} \left( \frac{\lambda^2}{c^2} + \frac{1}{4} \right) \right]
\end{equation}
and chemical potential $h$ adjusted to satisfy the filling condition $\int d\lambda \rho(\lambda) = n$.
This GTBA equation can amazingly be solved analytically in terms of modified Bessel functions of the first kind,
{ \setlength{\mathindent}{5mm}
\begin{eqnarray}
\rho_{sp} (\lambda) = -\frac{\gamma}{4\pi} \frac{1}{1 + a(\lambda)} \frac{\partial a(\lambda)}{\partial \gamma}, \hspace{5mm}
a(\lambda) \equiv \frac{2\pi}{\frac{\lambda}{n} \sinh \frac{2\pi \lambda}{c}} I_{1 - 2i\frac{\lambda}{c}} \left(\frac{4}{\sqrt{\gamma}}\right) I_{1 + 2i\frac{\lambda}{c}} \left(\frac{4}{\sqrt{\gamma}}\right).
\label{eq:BECrhosp}
\end{eqnarray}
}\noindent
As expected, the plot of these distributions (see Fig. 1 of \cite{2014_DeNardis_PRA_89}) shows an origin-centered, broadened peak which is non-thermal in shape. From this distribution, steady-state properties can be computed. In \cite{2014_DeNardis_PRA_89}, the local static density moments $g_{2,3}$ with 
\begin{equation}
g_l = \langle \rho_{sp} | \left[\psi^\dagger(0)\right]^l \psi(0)^l | \rho_{sp} \rangle/n^l
\end{equation}
were computed in the thermodynamic limit by using the above-obtained steady state and the method described in \cite{2011_Kormos_PRL_107,2011_Pozsgay_JSTAT_P01011}. The static structure factor 
\begin{equation}
S(x) = \langle \rho_{sp} | \rho(x) \rho(0) | \rho_{sp} \rangle
\end{equation}
was itself computed in a finite-size regularization of the steady state by using ABACUS \cite{2009_Caux_JMP_50} and an adaptation of the finite-temperature method developed in \cite{2014_Panfil_PRA_89}.

A direct, coordinate Bethe Ansatz-based numerical verification of the Quench Action predictions for the properties of the steady state was obtained in \cite{2015_Zill_PRA_91,2016_Zill_NJP_18}. Few-particle properties were well reproduced, 3-body ones less so, which is to be expected in view of finite-size corrections (the numerics was done for up to 5 particles; a 3-body term is then not a weak operator).

\subsubsection{Charges, divergences and GGE.}
Let us end this subsection with a reminder of the interesting history of the (attempts at a) solution of this quench protocol. The natural starting point, namely to attempt a GGE treatment, was pursued in \cite{arXiv12043889} based on the second-quantized representations of the conserved charges, the idea being that these can in principle be readily evaluated in the initial BEC state. 
For the Lieb-Liniger model, the natural charges to consider are those coming from the transfer matrix with evaluation point around infinity (see \cite{KorepinBOOK} and references therein). This leads to a hierarchy of charges $\hat{Q}^{(n)}$ with eigenvalue equations
\begin{equation}
\hat{Q}^{(n)} | \{ \lambda \}_N \rangle = Q^{(n)} (\{ \lambda \}_N), \hspace{5mm} Q^{(n)} (\{ \lambda \}_N) = \sum_{j=1}^N \lambda_j^n.
\end{equation}
The generic form for the GGE would then be, in operator and eigenvalue forms, 
\begin{equation}
\sum_{n=0}^\infty \beta_n \hat{Q}^{(n)}, \hspace{10mm} L \sum_n \beta_n \int d\lambda \rho(\lambda) \lambda^n,
\label{eq:BECGGE}
\end{equation}
with parameters $\beta_n$ determined from the initial conditions as per the GGE prescription. 

The implementation of the GGE for this problem however hits a snag. Higher conserved charges \cite{1990_Davies_PA_167,arXiv1109.6604} have terms which are not quantum mechanically normal-ordered. Although evaluating these charges (defined with $c$-dependent coefficients) on an eigenstate of the `correct' Hamiltonian (namely the same $c$ as that used in the definition of the charge) leads to a well-defined eigenvalue, due to mutual cancellations between delta functions originating from the kinetic and interaction terms. This cancellation however does {\it not} occur when evaluating a charge defined at $c'$ on a state defined at $c \neq c'$. 

This is reflected in the divergence of the $t=0$ expectation value of the higher charges on the BEC state, as can be seen from direct calculations \cite{CauxMosselBECunpublished}. The exact Quench Action solution for the steady state reproduces these divergences correctly: for all post-quench interaction values $c > 0$, the saddle-point distribution (\ref{eq:BECrhosp}) has a $1/\lambda^4$ tail at large rapidity,
\begin{equation}
\rho_{sp}(\lambda) = \frac{1}{2\pi} \left[ \frac{n^4 \gamma^2}{\lambda^4} + \frac{n^6 \gamma^3 (24 - \gamma)}{4\lambda^6} + ... \right]
\end{equation}
(these first two terms in the expansion were obtained in \cite{2013_Kormos_PRB_88} using q-bosons; the Quench Action solution can be expanded to any order if one feels motivated). The presence of this tail has dramatic consequences as far as conserved charges are concerned, since now all even charges with index $\geq 4$ become infinite, $\int \rho_{sp} (\lambda) \lambda^{2n} \rightarrow \infty$, $n \geq 2$. The divergence of $Q^{(4)}$ is like $\delta (x=0)$ (namely: the dimension of momentum space, infinite here since we are in the continuum without a UV cutoff), with higher charges diverging more and more strongly. 

The GGE logic therefore cannot be applied to this particular quench problem (which is another reason, at least for the author, to find this problem particularly interesting). These presence of these divergences motivated the study in \cite{2013_Kormos_PRB_88} (updating \cite{arXiv12043889}) by considering a regularization of the problem in terms of $q$-bosons, for which a partial GGE could be consistently implemented. A full GGE was however out of reach, and it remained for the Quench Action to provide for a full solution.

This problem with divergences is not confined to the BEC to Lieb-Liniger quench. The three preconditions for its existence are that the Hilbert space be infinite-dimensional ({\it e.g.} here on the continuum interval, there is no UV cutoff), that the maximal value of the interaction potential be unbounded (here, a delta function), and that the quench protocol involve a change of this interaction potential. It is thus anticipated that a more generic Lieb-Liniger quench protocol going from $c$ to a different $c'$ would display the same divergences, and also any form in interaction quenches in multi-species generalizations of Lieb-Liniger.

Using the perspective offered by the exact solution of the problem using the Quench Action, we can note that the forms (\ref{eq:BECGGE}) and (\ref{eq:BECoverlapsextensive}) are incompatible due to the logarithmic singularity in the latter (this being an exact result), which cannot be recaptured by power-sum (or polynomial-type) conserved charge eigenvalues. One suggestive way to `repair' the GGE for the BEC to Lieb-Liniger quench is thus to include a charge which does not directly originate from the usual trace identities coming from the transfer matrix. If one included a `log' charge with eigenvalue
\begin{equation}
Q^{\mbox{log}} (\{ \lambda_j \}) = \sum_{j=1}^N \ln \left[\lambda_j^2(\lambda_j^2 + (c/2)^2)\right],
\end{equation}
then the exact QA free energy (\ref{eq:BECoverlapsextensive}) trivially becomes of GGE form (with only this single charge!). This charge (a kind of `logarithm of particle energy') does not appear physically meaningful at first sight, though it is mathematically completely well defined (its matrix elements on the basis of Bethe states are all well-defined, at least for the even particle number sector) and it gives meaningful, extensive values on `normal-looking' Bethe states and is thus probably of (quasi?)local nature. This way of `repairing' the GGE does not correspond to a kind of UV regularization as offered by the q-boson approach. Whether one can ascribe any meaning to all of this is an open issue.

A physically perhaps more intuitive way to save the situation is to consider a slightly modified quench problem, whereby the initial state is not the perfect BEC state, but a UV/high-energy regularized one $e^{-\epsilon H_{LL}} | \Psi_0\rangle$. This modification of the initial state leads to the exponential suppression of the mode occupation at rapidities $\lambda \gg \frac{1}{\sqrt{\epsilon}}$, and thus to the regularization of all infinities in the initial values of the charges. This approach, which is also a kind of UV regularization, is discussed in more detail in \cite{FiorettoBECreg}.

\subsubsection{Time evolution.}
The time evolution following the BEC to Lieb-Liniger quench can in principle be computed using the Quench Action method, 
using the fundamental representation ({\ref{eq:tdepOsp}). Besides the saddle-point distribution (\ref{eq:BECrhosp}), one needs the differential overlap function (\ref{eq:dSexc}), which was calculated in the original paper \cite{2014_DeNardis_PRA_89} (see equation (A14)) and indeed obeys the expected decomposition (\ref{eq:overlapfn1}) as a sum of one-body terms,
\begin{equation}
\delta S_{\bf e} [\rho_{sp}] = \sum_{k=1}^n \left[ \delta s (\tilde{\lambda}_k^p) - \delta s (\tilde{\lambda}_k^{h}) \right],
\label{eq:BECdiffoverlaps}
\end{equation}
where $\tilde{\lambda}^{p,h}$ are the particle and hole excitation rapidities and
{ \setlength{\mathindent}{5mm}
\begin{equation}
\delta s (\tilde{\lambda}^p) - \delta s (\tilde{\lambda}^{h}) = \int_0^\infty d\lambda \rho_{sp}(\lambda) \frac{1 + 8 \frac{\lambda^2}{c^2}}{\lambda (1 + 4\frac{\lambda^2}{c^2})} F(\lambda | \tilde{\lambda}^p, \tilde{\lambda}^h) + \ln \left( \frac{\tilde{\lambda}^p \sqrt{(\tilde{\lambda}^p/c)^2 + 1/4}}{\tilde{\lambda}^h \sqrt{(\tilde{\lambda}^h/c)^2 + 1/4}} \right)
\end{equation}
}\noindent
with backflow function obeying 
{ \setlength{\mathindent}{5mm}
\begin{equation}
2\pi F (\lambda | \tilde{\lambda}^p, \tilde{\lambda}^h ) - \int_{-\infty}^\infty d\lambda' K (\lambda - \lambda') \vartheta_{sp} (\lambda') F(\lambda' | \tilde{\lambda}^p, \tilde{\lambda}^h ) = \phi (\lambda - \tilde{\lambda}^p) - \phi (\lambda - \tilde{\lambda}^h)
\end{equation}
}\noindent
in which $\vartheta_{sp} (\lambda) = [1 + \rho^h_{sp}(\lambda)/\rho_{sp}(\lambda)]^{-1}$ is the saddle-point state's filling function.

These ingredients were used in the specific case of a quench from the BEC state to hard-core (Tonks-Girardeau) bosons, to obtain the exact analytical expression for the time-evolved one-body density matrix \cite{2014_DeNardis_JSTAT_P12012}, this taking the form of a difference between two Fredholm Pfaffians. In addition, the previously-derived time-dependent density-density correlation \cite{2014_Kormos_PRA_89} 
was reobtained using the Quench Action logic. 

The QA was further employed to study the time dependence of observables at generic interaction in \cite{2015_De_Nardis_JPA_48}, suggesting that power-law relaxation should be generic at late times for observables which do not create large numbers of particle-hole excitations. Further investigation of specific examples of other observables is a promising research direction for the future.

\subsection{The N{\'e}el to XXZ quench}
Following up on the solution to the BEC to Lieb-Liniger quench described above, the Quench Action was then applied to a problem in the context of spin chains. More precisely, the initial state defined by the N{\'e}el state
\begin{equation}
| \Psi_0 \rangle \equiv \frac{1}{\sqrt{2}} \left( | \uparrow \downarrow \uparrow \downarrow ... \rangle + | \downarrow \uparrow \downarrow \uparrow ... \rangle \right)
\label{eq:Neel}
\end{equation}
was time-evolved using the $XXZ$ Hamiltonian 
\begin{equation}
H = J \sum_{j=1}^N \left[ S^x_j S^x_{j+1} + S^y_j S^y_{j+1} + \Delta S^z_j S^z_{j+1} \right]
\label{eq:XXZ}
\end{equation}
with anisotropy parameter $\Delta \geq 1$ lying in the antiferromagnetic region. The initial state (\ref{eq:Neel}) is the ground state of the Ising Hamiltonian at $\Delta \rightarrow \infty$, so this can be viewed as an `interaction' quench in the spin chain language. For definiteness, periodic boundary conditions are used.

In contrast to the repulsive Lieb-Liniger model, the $XXZ$ chain admits bound states, taking the form of string patters in the solution to the Bethe equations, 
\begin{equation}
\lambda_\alpha^{j, a} = \lambda_\alpha^j + i\frac{\eta}{2} (n+1 - 2a) + i \delta_\alpha^{j,a}, \hspace{10mm} a = 1, ..., n,
\label{eq:XXX:strings}
\end{equation}
 $\eta$ being related to the chain's anisotropy parameter by the relation $\Delta = \cosh \eta$. 
The real parameter $\lambda_\alpha^j$ represents the {\it string center}, namely the `center of mass' of the composite object represented by the $n$ rapidities. The string deviations $\delta_\alpha^{j,a}$ are in most circumstances exponentially small in system size, and can typically be neglected, in which case one works in the so-called string hypothesis. Under this, the Bethe equations become equations for the string centers, the Bethe-Gaudin-Takahashi (BGT) equations \cite{1971_Takahashi_PTP_46,1972_Takahashi_PTP_48,GaudinBOOK,TakahashiBOOK}
\begin{equation} 
\theta_j (\lambda^j_\alpha) - \frac{1}{N} \sum_{k=1}^{N_s} \sum_{\beta = 1}^{M_k} 
\Theta_{jk} (\lambda^j_\alpha - \lambda^k_\beta)
= \frac{2\pi}{N} I^j_\alpha
\label{eq:BAEXXZstrings}
\end{equation}
in which
\begin{equation}
\theta_j(\lambda) = 2\mbox{atan} \frac{\tan(\lambda)}{\tanh(n\eta/2)} + 2\pi \left\lfloor \frac{\lambda}{\pi} + \frac{1}{2} \right\rfloor
\end{equation}
and the string-string scattering phase shift is
{ \setlength{\mathindent}{10mm}
\begin{equation}
\Theta_{jk} (\lambda) = (1 - \delta_{jk}) \theta_{|j-k|} (\lambda)
+ 2\theta_{|j-k|+2} (\lambda) + ... + 2\theta_{j+k-2} (\lambda) + \theta_{j+k} (\lambda).
\label{eq:stringstringscat}
\end{equation}
}\noindent
Unlike in the Lieb-Liniger model (where there is no UV cutoff and thus no limit on the quantum numbers), there exist limiting quantum numbers at each string level, which depend on the fillings of each level (we omit those details here for the sake of brevity). The string center rapidities are defined in the interval $[-\frac{\pi}{2}, \frac{\pi}{2}]$. 

The thermodynamic limit of the XXZ chain is expressed as before in terms of densities, this time each string type having its own density function. For example, the expressions for the energy and magnetization are
{ \setlength{\mathindent}{10mm}
\begin{equation}
\frac{E}{N} = -\frac{h}{2} + \sum_{n=1}^\infty \int_{-\infty}^\infty d\lambda ~d_n (\lambda) \rho_n (\lambda), \hspace{5mm}
\frac{S^z_{tot}}{N} = \frac{1}{2} - \sum_{n=1}^\infty \int_{-\infty}^\infty d\lambda ~n \rho_n (\lambda)
\end{equation}
}\noindent
where the thermal driving term is $d_n (\lambda) = \beta(hn - \pi J a_n (\lambda))$, $a_n (\lambda) = \frac{1}{2\pi} \frac{d}{d\lambda} \theta_n (\lambda)$, while the expression for the Yang-Yang entropy is
\begin{equation}
S^{YY} [\{ \rho_n \}]/N = \sum_{n=1}^\infty \int_{-\pi/2}^{\pi/2} d\lambda \left[ \rho_n^t \ln \rho_n^t - \rho_n \ln \rho_n - \rho_n^h \ln \rho_n^h \right].
\end{equation}
The usual TBA thermodynamic equilibrium equations, obtained by the saddle-point evaluation of the partition function's functional integral representation, take the form
\begin{equation}
\ln \eta_n (\lambda) = \beta d_n(\lambda) + \sum_{m=1}^\infty A_{nm} * \ln (1 + \eta_m^{-1} (\lambda))
\label{eq:XXXTBA1}
\end{equation}
in which $A_{nm} (\lambda) = \frac{1}{2\pi} \frac{d}{d\lambda} \Theta_{nm} (\lambda)$ and
$\eta_n (\lambda) \equiv \rho_n^h (\lambda)/\rho_n (\lambda)$.

\subsubsection{Overlaps.}
As before, the necessary starting point when aiming to implement a Quench Action treatment is to study the overlaps between the initial state (\ref{eq:Neel}) and the exact eigenstates of (\ref{eq:XXZ}). Building on earlier work \cite{1998_Tsuchiya_JMP_39,2012_Kozlowski_JSTAT_P05021} (see also \cite{2014_Pozsgay_JSTAT_P06011}), it was shown in \cite{2014_Brockmann_JPA_47a,2014_Brockmann_JPA_47b} that these overlaps are given by \cite{2014_Brockmann_JPA_47a}
{ \setlength{\mathindent}{5mm}
\begin{equation}
  \frac{\langle \Psi_0 | \{\pm\lambda_j\}_{j=1}^{M/2} \rangle }{\|\{\pm\lambda_j\}_{j=1}^{M/2}\|}=\sqrt{2} \left[\prod_{j=1}^{M/2}\frac{\sqrt{\tan(\lambda_j+i\eta/2) \tan(\lambda_j-i\eta/2)}}{2\sin(2\lambda_j)}\right]\sqrt{ \frac{\det_{M/2}(G_{jk}^{+})}{\det_{M/2}(G_{jk}^{-})}}
\label{eq:Neeloverlap}
\end{equation}
}\noindent
where
{ \setlength{\mathindent}{5mm}
\begin{equation}
G_{jk}^\pm = \delta_{jk}\left(NK_{\eta/2}(\lambda_j)-\sum_{l=1}^{M/2}K_\eta^+(\lambda_j,\lambda_l)\right) + K_\eta^\pm(\lambda_j,\lambda_k)\:, \quad j,k=1,\ldots,M/2
\label{eq:overlap_b}
\end{equation}
}\noindent 
in which $K_\eta^\pm(\lambda,\mu)=K_\eta(\lambda-\mu) \pm K_\eta(\lambda+\mu)$, and $K_\eta(\lambda)=\frac{\sinh(2\eta)}{\sin(\lambda+i\eta)\sin(\lambda-i\eta)}$. Note that a `doubled' Gaudin matrix once again appears, this being reminiscent of the BEC overlaps (\ref{eq:BECoverlap}). This is natural in view of the fact that the latter can in fact be viewed as a limit of the former (see \cite{2014_Brockmann_JSTAT_P05006}). Also, once again, only parity-invariant states contribute.

\subsubsection{The steady state from the Quench Action.}
These ingredients were used in \cite{2014_Wouters_PRL_113} to implement the Quench Action protocol and obtain an exact solution to the N{\'e}el to XXZ quench problem (an extended treatment is published in \cite{2014_Brockmann_JSTAT_P12009}). Let us recall a few of the important aspects of this solution.

The extensive part of the overlaps are extracted from (\ref{eq:Neeloverlap}), verifying that string deviations do not ruin the limit. These give us the overlap functional (\ref{eq:So}) as
\begin{equation}
S^o[\{ \rho \}] = -\frac{N}{2} \sum_{n=1}^\infty \int_0^{\pi/2} d\lambda \rho_n (\lambda) \ln W_n (\lambda)
\end{equation}
with $W_n(\lambda)$ being rapidity- and anisotropy-dependent functions explicitly given in \cite{2014_Wouters_PRL_113}, equations (S7) and (S8).

Everything is then in position to enforce the saddle-point condition (\ref{eq:rhosp}), now generalized to many particle (string) types labeled by index $n$,
\begin{equation}
\left. \frac{\delta S^Q [\{\rho \}]}{\partial \rho_n (\lambda)}\right|_{\{ \rho_n \} = \{ \rho_{n,sp} \}} = 0,
\end{equation}
yielding \cite{2014_Wouters_PRL_113} equations similar to the thermal equilibrium ones (\ref{eq:XXXTBA1}) but with different driving terms (h being again a chemical potential/magnetic field enforcing the filling constraint of zero magnetization):
\begin{equation}
\ln \eta_n (\lambda) = 2n (\ln 4 - h) + g_n(\lambda) + \sum_{m=1}^\infty A_{nm} * \ln (1 + \eta_m^{-1} (\lambda)),
\label{eq:NeelXXZsp}
\end{equation}
with drivings again given by logarithms,
\begin{equation}
g_n (\lambda) = \sum_{l=0}^{n-1} \ln \left[ \frac{\sin^2 (2\lambda) + \sinh^2 [\eta(n-1-2l)]}{4 |\tan[\lambda + i \frac{\eta}{2}(n - 2l)]|^2} \right].
\end{equation}
Similarly to the thermal equilibrium case, there exists also a more aesthetic form of these coupled equations making use of Takahashi's decoupling scheme. We omit these here for succinctness and refer to \cite{2014_Wouters_PRL_113,2014_Brockmann_JSTAT_P12009} for this alternative version.

\subsubsection{GGE using the transfer matrix charges.}
A GGE treatment including the traditional charges (those emanating from the usual `spin-1/2' transfer matrix of the form (\ref{eq:tauistraceT}), quasilocal charges not being known at the time) was proposed independently in \cite{2013_Pozsgay_JSTAT_P07003} and \cite{2013_Fagotti_JSTAT_P07012} (the results of the latter subsequently extended in \cite{2014_Fagotti_PRB_89}). 
The GGE was thereafter exactly implemented using all these charges in \cite{2014_Wouters_PRL_113}. 
A rather surprising fact was obtained in \cite{2014_Wouters_PRL_113,2014_Brockmann_JSTAT_P12009}, namely that the constraints 
\begin{equation}
\lim_{t\rightarrow \infty} \langle \Psi (t) | \hat{Q}^{(a)} | \Psi(t) \rangle = \langle \{ \rho_{n,sp} \} | \hat{Q}^{(a)} | \{ \rho_{n,sp} \} \rangle
\end{equation}
do not fix all the densities of the steady state, but rather only fix the density of holes of one-strings,
\begin{equation}
\rho^h_{1; sp} (\lambda) = \frac{\pi^2 a_1^3 (\lambda) \sin^2(2\lambda)}{\pi^2 a_1^2 (\lambda) \sin^2(2\lambda) + \cosh^2(\eta)}.
\label{eq:rho1hNeel}
\end{equation}
This means that all other densities corresponding to higher strings are actually left free to fluctuate, meaning that constraints seem to be missing, as also noticed in \cite{2014_Goldstein_PRA_90,2014_Pozsgay_JSTAT_P09026}.

This fact is indeed demonstrated by comparison to the exact Quench Action solution to the problem, leading to the somewhat shocking and unexpected conclusion that the GGE based on these traditional charges only, failed to converge to the exact Quench Action prediction for the steady state, the difference being markedly illustrated in the different distributions of even-length bound states at the origin (see Fig. 1 in \cite{2014_Wouters_PRL_113}). These differences in distributions affect the observables, for example the local spin correlations; these differences were verified using numerics based on a linked-cluster expansion. In a back-to-back paper \cite{2014_Pozsgay_PRL_113} (with extended version \cite{2015_Mestyan_JSTAT_P04001}), the results for the N{\'e}el quench were reproduced and extended to the dimer product state
{ \setlength{\mathindent}{5mm}
\begin{equation}
|\Psi (t=0) \rangle = \frac{1}{2^{(N+1)/2}} \Bigl( | (\uparrow \downarrow - \downarrow \uparrow) (\uparrow \downarrow - \downarrow \uparrow) ... \rangle + |(\downarrow \uparrow - \uparrow \downarrow) (\downarrow \uparrow - \uparrow \downarrow) ... \rangle \Bigr)
\end{equation}
}\noindent
(corresponding to the ground state of the Majumdar-Ghosh model \cite{1969_Majumdar_JMP_10}).
The great virtue of the dimer state is that it showed an even more dramatic contrast between the Quench Action and GGE predictions than the N{\'e}el case, with the discrepancies being clearly exposed in numerical verifications using infinite size time evolved block decimation. This study was later extended to include a wider class of q-dimer initial states (for which exact overlaps are known \cite{2014_Pozsgay_JSTAT_P06011}) in \cite{2015_Mestyan_JSTAT_P04001}, including more a detailed study of the local longitudinal and transverse spin-spin correlations. These studies provided additional confirmation that the Quench Action provided the correct answers for the steady state.

As far as the steady state is concerned, the story concerning the discrepancy between the exact Quench Action results and the GGE was brought to a close with the understanding that the correct GGE \cite{2015_Ilievski_PRL_115_15} in the case of quantum spin chains needed to also involve families of quasilocal conserved charges generated from higher-spin transfer matrices. We refer to the accompanying paper \cite{arXiv160300440} for an extensive introduction to these quasilocal charges and to their use in different contexts, and to \cite{arXiv151203713} for a rigorous discussion of locality as far as thermalization is concerned. We however point out that these quasilocal charges, first derived in \cite{2015_Ilievski_PRL_115_12}, can be used when quenching to the gapless regime \cite{arXiv1512.04454}, this representing another road to obtain the steady state (but not the time evolution).

\subsection{Other applications}
To close this section, although it is not our intent to review the field here, let us nonetheless briefly mention that the Quench Action has already been used in many other contexts in addition to those already described, including the sine-Gordon \cite{2014_Bertini_JSTAT_P10035} and sinh-Gordon \cite{arXiv1602.08269} models, geometric quenches in free fermionic chains \cite{2015_DeLuca_PRA_91}, finite integrable spin chains \cite{2016_Alba_JSTAT_043105}, the Kondo model \cite{2015_Bettelheim_JPA_48}, the Lieb-Liniger gas with hard-wall boundary conditions \cite{2015_Goldstein_PRB_92}, Bragg pulses in bosonic gases \cite{arXiv150706339}, quenches from rotating BECs \cite{arXiv1510.08125}, etc. Another very interesting case is that of the extension of the BEC to repulsive Lieb-Liniger quench work \cite{2014_DeNardis_PRA_89} to the {\it attractive} case \cite{2016_Piroli_PRL_116}, which also relates to the KPZ equation \cite{2014_Calabrese_JSTAT_P05004}. Since this model supports bound state excitations, the implementation of the Quench Action requires treating multiple root distributions, similarly to what needs to be done in the case of quantum spin chains. These illustrate the variety of contexts in which the method can be fruitfully applied.

\section{Roadmap and perspectives}
\label{sec:Road}
\subsection{The hunt for overlaps}
Taking a step away from specific cases of implementation of the Quench Action, it is worthwhile discussing the most important building blocks of the Quench Action, namely the exact overlaps which stand at the base of the whole formalism. Restricting our discussion for the moment to the case of quantum quenches, the general problem is to find workable expressions for the overlaps 
\begin{equation}
\langle \Psi_0 | \{ I \} \rangle_{g} \equiv \langle \Psi_0 | \{ \lambda \} \rangle_g
\end{equation}
between some initial state $| \Psi \rangle$ and the eigenstates of a certain (preferably, though not necessarily integrable) Hamiltonian with interaction parameter $g$, which we have denoted alternately in terms of quantum numbers or rapidities solving the relevant Bethe equations (including normalization).

The two cases we have discussed in more detail, namely the BEC to Lieb-Liniger quench on the one hand, and the N{\'e}el to XXZ quench in the other, rely on overlaps described in \cite{1998_Tsuchiya_JMP_39,2012_Kozlowski_JSTAT_P05021,2014_Brockmann_JPA_47a,2014_Brockmann_JPA_47b}. Other known results include overlaps between states at $\Delta = \infty$ and $\Delta = 0$ \cite{2016_Mazza_JSTAT_P013104} and of partial N{\'e}el states \cite{2016_Foda_JSTAT_P023107}. An interesting further proposal provides some recursive expressions for overlaps of simple product states \cite{2014_Piroli_JPA_47}. Perhaps one could extend these all the way to any initial state satisfying cluster decomposition, these being representative of the generic case (see for example (\cite{2014_Sotiriadis_JSTAT_P07024}).

It is however not trivial to generalize these results to more general initial states. In fact, these two problems share many similarities with each other. The resemblance between (\ref{eq:BECoverlap}) and (\ref{eq:Neeloverlap}) is particularly striking, and we can wonder how general similar constructions could be. Gaudin's formula, giving the norm of a Bethe eigenstate in terms of the determinant of the Gaudin matrix, has a clear intuitive origin: the quantum numbers of an eigenstate can be viewed as the proper independent variables labelling a state, and the measure for summation over eigenstates is flat in quantum number space. Since the Bethe equations define the mapping between quantum number space and rapidity space, the measure for summation over eigenstates, when translated to rapidity space, gets rescaled by the Jacobian of the transformation between quantum numbers and rapidities, namely by the determinant of the Gaudin matrix whose entries are the derivatives of the Bethe equations (second derivatives of the Yang-Yang action). The fact that the overlaps (\ref{eq:BECoverlap}) and (\ref{eq:Neeloverlap}) are given by a Gaudin-like matrix is however less intuitive from the outset: these overlaps can be viewed as a kind of `square-root norm' originating from the partition function of the 6-vertex model with reflecting boundary conditions \cite{1998_Tsuchiya_JMP_39}, but the deeper meaning of these identities remains obscure at this stage.

An interesting open challenge is to try to generalize (\ref{eq:BECoverlap}) to the case of a generic free boson initial state 
\begin{equation}
| \Psi_0 \rangle = | \{ k \}_N \rangle.
\label{eq:genfreebosoninitial}
\end{equation}
For generic choices of momenta $\{ k \}_N$, the overlap with Bethe states cannot be of Gaudin-like form (\ref{eq:BECoverlap}). In fact, the perspective for the obtention of a genuinely useful representation of the overlaps is pessimistic: one can create the states (\ref{eq:genfreebosoninitial}) by applying a product of single-particle creation operators on the vacuum; alternately, the overlap can be viewed as the overlap of the vacuum with the remnants of the Bethe states after acting with the free boson annihilation operators. Whether the in principle factorially large sum which remains can be expressed in an economical determinant-like form remains an open issue at this stage. Note however that the obtention of a workable formula here would in principle solve the extremely interesting and important problem of calculating overlaps of eigenstates at different values of the interaction parameter,
\begin{equation}
{}_g \langle \{ I \} | \{ I \} \rangle_{g'},
\end{equation}
by simply using the free bosonic basis as an intermediate basis. We leave this problem as an outstanding challenge to the community, noting in passing that there is one case in which such a formula can be obtained: for the Richardson model
\begin{equation}
H = \sum_\alpha \sum_{\sigma = \uparrow, \downarrow} \frac{\varepsilon_\alpha}{2} c^\dagger_{\alpha \sigma} c_{\alpha \sigma} - g \sum_{\alpha_1, \alpha_2} c^\dagger_{\alpha_1 \uparrow} c^\dagger_{\alpha_1 \downarrow} c_{\alpha_2 \downarrow} c_{\alpha_2 \uparrow},
\end{equation}
one can write the exact overlap between states at different values of $g$  \cite{2009_Faribault_JSTAT_P03018} by invoking Slavnov's theorem. Although the Quench Action method is not usable here (the thermodynamic limit of this model is inherently mean-field), the algebraic structure of the overlaps might serve as inspiration for other non mean-field-like cases.

The problem we just mentioned, namely that of the overlap between exact eigenstates of Hamiltonians with differing interaction parameters, is formally written in terms of Algebraic Bethe Ansatz operators as
\begin{equation}
\langle 0 | \prod_{j} C^{(g)} (\lambda^{(g)}_j) \prod_k B^{(g')} (\mu^{(g')}_{k}) | 0 \rangle
\end{equation}
in which we have explicitly labeled the monodromy matrix operators $B, C$ in terms of the interaction parameter since the algebra of these operators explicitly depends on these interaction parameters. Similarly, the sets of rapidities are superscripted by the value of the interaction parameter for which they solve Bethe equations. Although the commutation relations between ABA operators pertaining to a given interaction parameter are relatively simple (namely the usual quadratic algebra), the `cross'-algebra giving the commutation relations between operators at different interaction values is dramatically more complicated. One can however dream of a higher algebraic structure intertwining the ABA operators at $g$ with those at $g'$; this structure, which might take the form of some exponentiated, continuous unitary transformation, remains however elusive at this stage. A more immediately workable but still interesting possibility is thus to exploit the crucial feature making Slavnov's theorem interesting, namely that the overlaps
\begin{equation}
\langle 0 | \prod_{j} C (\lambda_j) \prod_k B (\mu_{k}) | 0 \rangle
\end{equation}
(in which we now consider a single interaction parameter $g$) are exactly known provided {\it one} of the sets $\{ \lambda \}$ or $\{ \mu \}$ solves the Bethe equations, the other set being completely arbitrary. It is thus possible to immediately write ovelaps between Bethe states and generic initial states 
\begin{equation}
| \Psi_0 \rangle \equiv | \{ \mu \} \rangle
\label{eq:Slavnovinit}
\end{equation}
for arbitrary (but judiciously chosen so at to make the problem interesting) state-defining rapidity set $\{ \mu \}$. Although the precise nature of such states is not immediately obvious, it can in principle be quantified by mapping the states back to Bethe states (again invoking Slavnov's theorem) and computing representative observables. By cleverly choosing these sets of rapidities based on known intuitions, one can define whole families of quench problems whose solution via the Quench Action becomes straightforward. We suggest this as a line worth exploring in the future.

Finally, an important point to bear in mind for the application of the Quench Action method is that the knowledge of the exact overlaps to all orders in system size is overkill, at least as far as the determination of the steady state is concerned. For this purpose, only the leading extensive part of the logarithmic overlap is needed. It might be possible to devise simplified schemes in which such leading parts can be obtained, bypassing exact calculations. Results on these, when fed back into the Quench Action formalism, would allow to reconstruct the full time evolution of observables from quench time to the steady state. We again leave this as an open challenge.

\subsection{Simplicity and solvability}
In both the BEC to Lieb-Liniger and the N{\'e}el to XXZ quenches, remarkably, the simplicity of the original state has imprinted itself in the analyticity of the solution to the GTBA for the steady state. 
The existence of a closed-form solutions for (generalized) TBA equations is atypical (an interesting early example being the appearance of Airy functions in integrable $N=2$ supersymmetric models \cite{1999_Fendley_LMP_49}). 
What the deeper meaning of this fact is, and whether such a `simplicity-analyticity' correspondence holds for other quenches, are interesting questions for the future.

\subsection{Relaxation: it's not the destination, it's the journey}
The one direction which probably offers the most prospects for new results is the one of the analysis of the time-dependent behaviour of observables after quenches. It is natural for theoretical efforts to have been up to now mostly directed at the characterization of steady states, since off-diagonal terms in the expectation values can then be dropped. On the other hand, the presence of any integrability-breaking term \cite{2014_Essler_PRB_89,2015_Bertini_PRL_115,2015_Bertini_JSTAT_P07012,2015_Brandino_PRX_5,2015_Konstantinidis_PRE_91}, even arbitrarily weak, destroys the long-time limit, presumably thermalizing it, though the timescales for this can be very large. Shifting the focus away from the steady state, it is obvious that even more interesting physics is to be obtained in the actual relaxation process itself, for which the Quench Action offers a handle. We have showed some results on this in the case of the BEC quench in Lieb-Liniger. These constructions could be extended to more complicated observables at the cost of surmountable computational difficulties. In the Tonks-Girardeau limit, the explicit time dependence can be constructed (from direct calculations which are easily reproducible from the Quench Action, see \cite{arXiv150706339}) for quenches such as the trap release \cite{2013_Collura_PRL_110,2013_Collura_JSTAT_P09025} (also to a gas in a hard-wall trap \cite{2016_Mazza_JSTAT_P11016}) or a Bragg pulse \cite{arXiv150706339}. In the low-density limit, generic quenches also lead to Tonks-Girardeau behaviour \cite{2013_Iyer_PRA_87}.

Time dependence in the case of the N{\'e}el quench or of the dimer quench in spin chains (studied in {\it e.e.g} \cite{2014_Fagotti_PRB_89}) is something which is in principle accessible to the Quench Action but remains to be done. At generic interaction value, for complicated but meaningful quantities such as (relative) time- and space-dependent spin correlations, one difficulty is that it then requires the computation of matrix elements on states with mixed string contents, in highly-excited quantum number configurations. This has been explored in some recent works \cite{2011_Pozsgay_JSTAT_P01011,2014_Mestyan_JSTAT_P09020,2015_DeNardis_JSTAT_P02019}. Numerically, it is possible to perform the necessary summations in (\ref{eq:tdepOsp}) by using the ABACUS algorithm, whose logic is applicable starting from any (even highly excited) state, though the summations become difficult to perform due to the finite entropy of the starting state and to the large system sizes required.

\subsection{Towards a phenomenological classification of relaxation behaviours}
We have seen before that the implementation of the Quench Action requires as ingredients: the saddle-point distribution $\rho_{sp}$ and the characteristic quench overlap function $s$. Given the excitation dispersion relations $\varepsilon$ (defined by $\rho_{sp}$ if the Hamiltonian is specified), and given an observable ${\cal O}$ which we are interested in, a knowledge of its matrix elements then allows to compute the post-quench time dependent expectation value, using the fundamental representation (\ref{eq:tdepOsp}), `working backwards' from the steady state by including more and more excitations.

Going further with the idea of `working backwards', one can simply turn the tables around and back-engineer quench problems by {\it postulating} the saddle-point $\rho_{sp}$ and characteristic overlap function $s$. Given these, one can then try to solve the {\it inverse} problem of determining which kind of state stood at the pre-quench origin of time. The level of exactitude with which this state can be determined depends on the detailed knowledge of $s$. Starting with the basics, one can imagine that many different initial states relax to the same saddle-point $\rho_{sp}$. How this relaxation occurs then becomes dependent on the interplay between the characteristics of $\rho_{sp}$ ({\it e.g.} its inflection points) and those of the overlap function $s$ and dispersion relation $\varepsilon$. One could thus develop the notion of `universality classes' of types of relaxation behaviour based on these characteristics.

\subsection{Driven systems}
An interesting exploratory route is to consider a much more brutally out-of-equilibrium situation, namely a periodically-driven system, in other words Floquet dynamics. The driving would then be periodic over a period $T$; in its simplest form, one can envision a `quench-dequench' cycle switching between Hamiltonians $H_1$ (for a duration of $t_1$) and $H_2$ (duration $t_2$, with $T = t_1 + t_2$) with Floquet operator
\begin{equation}
e^{-i t_2 H_2} e^{-i t_1 H_1}.
\end{equation}
Starting from a given initial state, repeated application of the Floquet operator will keep the system from equilibrating. If one focuses on observables measured at stroboscopic times (namely always at the same instant in the Floquet cycle), one can however expect pseudo-relaxation, in the sense that the system would tend to effectively relax to one of the set of Floquet eigenstates. The logic of the Quench Action can be applied to describe such stroboscopic observables, the Floquet eigenstates now taking on the previous role played by the Hamiltonian eigenstates in the quench problem. A steady-state can be expected (i.e. relaxation to a saddle-point Floquet eigenstate, see an example in spin chains in \cite{2000_Prosen_PTPS_139}) again due to the large relative dephasings. Finding this stroboscopic steady state would again be the result of an optimization procedure following the logic of the Quench Action. We have not yet been able to control such a calculation, and leave the finding of such an example as another open problem.

\subsection{Non-integrable models}
Although we have used the language of Bethe Ansatz to present the logic of the Quench Action, this is not a necessary requirement. Given a computationally useful basis of eigenstates, precisely the same steps can be followed in complete generality, the whole edifice being based on the sole existence of such an eigenbasis. The reason why Bethe Ansatz solvable models are productively used in actual implementations of the Quench Action is that they provide all the necessary ingredients: states and their quantum number labels, overlaps, matrix elements. One could imagine for example a future in which numerical methods, perhaps based on matrix-product states, are able to deliver a set of practical tools from which quench problems can be treated using the Quench Action logic. Admittedly, this is at the moment wishful thinking, but who knows what the future will bring.

\section{Conclusions}
In the short time since its formulation, the Quench Action has already shown itself to be a useful framework for computing the time-dependent expectation values of operators after quenching to an integrable model (alternately, after releasing an arbitrary initial state into such a system), valid for arbitrary times all the way to the (dephased and equilibrated) steady state. We have shown that in the thermodynamic limit, for a wide class of operators (those which are not entropy-producing), it allows to encode the full time evolution of expectation values into the very economical representation (\ref{eq:tdepOsp}). The existence of this representation relies on the fact that the only significantly contributing states are to be found in the vicinity of a `saddle point' state, obtained as the extremum state of an effective action functional, the Quench Action, which is fully specified by the initial conditions/quench protocol. Besides giving a clear mechanism for calculations in explicitly-defined protocols, the treatment presented here opens the door to a phenomenological classification of the possible time-dependent behaviours by `back-engineering' an inverse problem starting from the steady state.

Using the Quench Action framework, a number of exact results have been obtained for quenches to nontrivially interacting problems in the thermodynamic limit, some of which were summarized here. It is expected that many more such examples exist and can be found starting from the knowledge and techniques currently at our disposal.

The examples discussed in detail here were in translationally-invariant systems. The Quench Action has already been applied to situations where this invariance is broken, {\it e.g.} \cite{2015_DeLuca_PRA_91,2015_Bettelheim_JPA_48}. In the absence of translational invariance, interesting aspects emerge. Most simply, if the initial state is not translationally invariant, extra charges must then be included in the GGE \cite{2014_Fagotti_JSTAT_P03016,2015_Bertini_JSTAT_P07012,arXiv1507.02678} although the time evolution is driven by a translationally-invariant Hamiltonian. A different context one could consider in the future is that of many-body localization \cite{2015_Nandkishore_ARCMP_6} (see also the accompanying paper \cite{arXiv1603.06618}), where translational invariance is also severely broken. The links to relaxation in classical integrable systems (see the accompanying paper \cite{arXiv1603.08628}) can also be made.

Results from the Quench Action also provide extremely stringent tests for other methods. Notably, the exact solution for the N{\'e}el to XXZ quench using the Quench Action has enabled the correct formulation of the GGE for interacting models (at least in the context of spin chains), where extended sets of (quasilocal) charges were needed to reproduce the exact QA solution. In the context of interaction quenches in 1d gases, the QA solution to the BEC to Lieb-Liniger quench exists but has not yet been reproduced using a properly-formulated GGE. Looking further at beyond-GGE properties, the time dependence of observables (which the QA can also access, though the number of worked-out cases here is more limited), could similarly provide tough benchmarks for time-dependent numerical methods like tDMRG and ITEBD. The link with out-of-equilibrium conformal field theory (see the accompanying papers \cite{arXiv1603.02889,arXiv1603.07765}) and Luttinger physics (see \cite{arXiv1603.04252}) could also be explored further. Finally, these finite-time calculations could potentially be used to provide explicit experimental phenomenology, for example in the context of prethermalization (see the accompanying paper \cite{arXiv1603.09385}), extending what has been achieved for the equilibrium dynamics of integrable spin chains \cite{2009_Thielemann_PRL_102,2009_Walters_NATPHYS_5,2011_Rule_PRB_84,2012_Schlappa_NATURE_485,2013_Mourigal_NATPHYS_9,2013_Lake_PRL_111} and Bose gases \cite{2015_Fabbri_PRA_91,2015_Meinert_PRL_115} to an even richer out-of-equilibrium context.

\ack
The author acknowledges useful discussions with R. van den Berg, G. Brandino, M. Brockmann, J. De Nardis, I. S. Eli{\"e}ns, F. H. L. Essler, D. Fioretto, A. G. Green, E. Ilievski, R. M. Konik, G. Mussardo, M. Panfil, M. Rigol, R. Vlijm, B. Wouters, and thanks the KITP in Santa Barbara for hospitality. This work was supported by the Foundation for Fundamental Research on Matter (FOM) and the Netherlands Organisation for Scientific Research (NWO), and forms part of the activities of the Delta Institute for Theoretical Physics (D-ITP).

\vspace{5mm}

\bibliographystyle{unsrt}
\bibliography{/Users/jscaux/WORK/BIBTEX_LIBRARY/BIBTEX_LIBRARY_JSCaux_PAPERS.bib,/Users/jscaux/WORK/BIBTEX_LIBRARY/BIBTEX_LIBRARY_JSCaux_BOOKS.bib,/Users/jscaux/WORK/BIBTEX_LIBRARY/BIBTEX_LIBRARY_JSCaux_OTHERS.bib,/Users/jscaux/WORK/BIBTEX_LIBRARY/BIBTEX_LIBRARY_JSCaux_OWNPAPERS.bib}

\end{document}